\documentclass[11pt,a4paper,english,nofootinbib]{revtex4}
\usepackage{lmodern}
\usepackage{lmodern}

\usepackage[T1]{fontenc}
\usepackage[latin9]{inputenc}
\usepackage{babel}
\usepackage{mathrsfs}
\usepackage{amsmath}
\usepackage{amssymb}
\usepackage{esint}
\usepackage[unicode=true,pdfusetitle,
 bookmarks=true,bookmarksnumbered=false,bookmarksopen=false,
 breaklinks=false,pdfborder={0 0 1},backref=false,colorlinks=false]
 {hyperref}

\makeatletter

\@ifundefined{textcolor}{}
{%
 \definecolor{BLACK}{gray}{0}
 \definecolor{WHITE}{gray}{1}
 \definecolor{RED}{rgb}{1,0,0}
 \definecolor{GREEN}{rgb}{0,1,0}
 \definecolor{BLUE}{rgb}{0,0,1}
 \definecolor{CYAN}{cmyk}{1,0,0,0}
 \definecolor{MAGENTA}{cmyk}{0,1,0,0}
 \definecolor{YELLOW}{cmyk}{0,0,1,0}
}

\usepackage{babel}
\usepackage{babel}
\usepackage{babel}
\usepackage{babel}
\usepackage{babel}

\usepackage{babel}
\usepackage{graphicx}
\def\b{\begin{equation}}
\def\e{\end{equation}}

\@ifundefined{textcolor}{}{%
 \definecolor{BLACK}{gray}{0}
 \definecolor{WHITE}{gray}{1}
 \definecolor{RED}{rgb}{1,0,0}
 \definecolor{GREEN}{rgb}{0,1,0}
 \definecolor{BLUE}{rgb}{0,0,1}
 \definecolor{CYAN}{cmyk}{1,0,0,0}
 \definecolor{MAGENTA}{cmyk}{0,1,0,0}
 \definecolor{YELLOW}{cmyk}{0,0,1,0}
 }

\usepackage{latexsym}\usepackage{bm}

\makeatother

\begin{document}
\title{{\normalsize{}Second Order Gauge Invariant Perturbation Theory and
Conserved Charges in Cosmological Einstein's Gravity}}
\author{{\normalsize{}Emel Altas}}
\email{emelaltas@kmu.edu.tr}

\affiliation{Department of Physics,\\
 Karamanoglu Mehmetbey University, 70100, Karaman, Turkey}
\date{{\normalsize{}\today}}
\begin{abstract}
{\normalsize{}Recently a new approach in constructing the conserved
charges in cosmological Einstein's gravity was given. In this new
formulation, instead of using the explicit form of the field equations
a covariantly conserved rank four tensor was used. In the resulting
charge expression, instead of the first derivative of the metric perturbation,
the linearized Riemann tensor appears along with the derivative of
the background Killing vector fields. Here we give a detailed analysis
of the first order and the second order perturbation theory in a gauge-invariant
form in cosmological Einstein's gravity. The linearized Einstein tensor
is gauge-invariant at the first order but it is not so at the second
order, which complicates the discussion. This method depends on the
assumption that the first order metric perturbation can be decomposed
into gauge-variant and gauge-invariant parts and the gauge-variant
parts do not contribute to physical quantities.}{\normalsize\par}
\end{abstract}
\maketitle

\section{{\normalsize{}Introduction}}

In General Relativity finding an exact solution is often very difficult
and therefore one needs to use perturbation theory, by starting from
an exact background solution with symmetries, which provides a lot
of information about the physical problem at hand. In the absence
of a source, any generic gravity field equations in local coordinates
read
\begin{equation}
\mathscr{E}_{\mu\nu}(g(\lambda))=0,\label{genericfieldequation}
\end{equation}
here $\lambda$ parametrizes the solution set. We have the exact solution
plus the perturbations defined as
\begin{equation}
g(\lambda=0):=\bar{g},~~~~~~~~h_{\mu\nu}:=\frac{dg_{\mu\nu}}{d\lambda}\Bigl|_{\lambda=0},~~~~~~~~k_{\mu\nu}:=\frac{1}{2}\frac{d^{2}g_{\mu\nu}}{d\lambda^{2}}\Bigl|_{\lambda=0},\label{eq:definition}
\end{equation}
where $\bar{g}$ is the background solution that we carry out the
perturbations around, $h$ denotes the first order perturbation of
the metric tensor and $k$ denotes the second order perturbation.
When we consider the perturbation of the field equations (\ref{genericfieldequation})
about the background spacetime solution $\bar{g}$, we obtain expansion
of the field equations up to $\mathcal{O}(\lambda^{3})$ as
\begin{equation}
\bar{\mathscr{E}}_{\mu\nu}(\bar{g})+\lambda(\mathscr{E}_{\mu\nu})^{\left(1\right)}(h)+\lambda^{2}\bigg((\mathscr{E}_{\mu\nu})^{\left(2\right)}(h,h)+(\mathscr{E}_{\mu\nu})^{\left(1\right)}(k)\bigg)=0.\label{expansion1}
\end{equation}
Here by assumption $\bar{\mathscr{E}}_{\mu\nu}(\bar{g})=0$ and $(\mathscr{E}_{\mu\nu})^{\left(1\right)}(h)$
denotes the first order linearized field equations while the combination
$(\mathscr{E}_{\mu\nu})^{\left(2\right)}(h,h)+(\mathscr{E}_{\mu\nu})^{\left(1\right)}(k)$
denotes the second order perturbations of the field equations. Of
course not all background solutions can be integrable to an exact
solution, since once $\bar{g}$ solves the background field equations,
the solution of the first order linearized field equations, $h$,
must satisfy the given relation (\ref{eq:definition}). Similarly
the second order metric perturbation must satisfy the given definition
with the second order field equations

\begin{equation}
(\mathscr{E}_{\mu\nu})^{\left(2\right)}(h,h)+(\mathscr{E}_{\mu\nu})^{\left(1\right)}(k)=0.\label{secondorderperturbedfieldequations}
\end{equation}
It means even if we find the linearized solutions, $h$, to the first
order perturbations of the field equations $(\mathscr{E}_{\mu\nu})^{\left(1\right)}(h)=0$,
there exists an additional constraint on it which comes from the second
order field equations. To see this situation explicitly let us consider
$\bar{\xi}^{\mu}$, a Killing vector field of the background spacetime.
Contraction of (\ref{secondorderperturbedfieldequations}) with $\bar{\xi}^{\mu}$
and integration of the result over a hypersurface $\varSigma$ of
the spacetime manifold $\mathscr{M}$ gives
\begin{equation}
\intop_{\varSigma}d^{n-1}x\thinspace\sqrt{\bar{\gamma}}\thinspace\bar{\xi}_{\mu}\thinspace(\mathscr{E}^{\mu\nu})^{\left(1\right)}(k)=-\intop_{\varSigma}d^{n-1}x\thinspace\sqrt{\bar{\gamma}}\thinspace\bar{\xi}_{\mu}\thinspace(\mathscr{E}^{\mu\nu})^{\left(2\right)}(h,h),\label{bye}
\end{equation}
where we have used the background metric and the inverse metric to
lower and raise the indices respectively and $\bar{\gamma}$ denotes
the metric of the hypersurface. Once the field equations of the theory
are given, we can express the left-hand side of (\ref{bye}) as a
pure divergence of an antisymmetric field $F^{\mu\nu}$
\begin{equation}
\sqrt{\bar{\gamma}}\bar{\xi}_{\mu}\thinspace(\mathscr{E}^{\mu\nu})^{\left(1\right)}(k)=\partial_{\mu}\left(\sqrt{\bar{\gamma}}F^{\mu\nu}\right).
\end{equation}
When the left-hand side of (\ref{bye}) is expressed in terms of the
metric perturbation, it is known as the Abbott-Deser-Tekin (ADT) current
(or charges) \cite{AD,DT} and it is an extension of the Abbott-Deser-Misner
(ADM) \cite{ADM} charges of flat spacetime. Substituting the last
expression in (\ref{bye}) we conclude that the right-hand side, which
is called the Taub charge \cite{Taub}, must also be expressed as
a pure boundary. Then one ends up with the equality of the Taub and
ADT charges
\begin{equation}
Q_{ADT}:=\intop_{\partial\varSigma}d\varSigma_{\mu}\thinspace\sqrt{\bar{\sigma}}\thinspace\hat{n}_{\nu}\thinspace\bar{\xi}_{\mu}\thinspace F^{\mu\nu}=-\intop_{\varSigma}d^{n-1}x\thinspace\sqrt{\bar{\gamma}}\thinspace\bar{\xi}_{\mu}\thinspace(\mathscr{E}^{\mu\nu})^{\left(2\right)}(h,h)=:-Q_{Taub},\label{equalityofcharges}
\end{equation}
where $\partial\varSigma$ is the boundary of the hypersurface $\varSigma$,
$\bar{\sigma}$ is the pull-back metric on it and $\hat{n}_{\nu}$
is the outward unit normal vector on $\partial\varSigma$. If the
background spacetime has no boundary, one arrives at the integral
constraint on the solutions of the linearized equations
\begin{equation}
\boxed{\phantom{\frac{\frac{\xi}{\xi}}{\frac{\xi}{\xi}}}\intop_{\varSigma}d^{n-1}x\thinspace\sqrt{\bar{\gamma}}\thinspace\bar{\xi}_{\mu}\thinspace(\mathscr{E}^{\mu\nu})^{\left(2\right)}(h,h)=0.}\label{integralconstraint}
\end{equation}
 When this integral constraint is satisfied, we say $\bar{g}$ is
linearization stable and the perturbation $h$ can be integrable to
an exact solution, but if this is not the case the background solution
has linearization instability and we cannot improve it to get an exact
solution, in other words $\bar{g}$ is an isolated solution. This
issue was studied for Einstein's theory in \cite{Deser-Brill,Deser-Bruhat,Fischer-Marsden,Fischer-Marsden-Moncrief,Marsden,Moncrief,Arms-Marsden},
summarized in \cite{Bruhat,Girbau-Bruna}; and it was extended to
the generic gravity theories recently in \cite{altasuzunmakale,emeltez}
and to chiral gravity in \cite{emelchiral}. For the cosmological
Einstein's theory it was shown that $\sqrt{\bar{\gamma}}\thinspace\bar{\xi}_{\mu}\thinspace(\mathscr{E}^{\mu\nu})^{\left(2\right)}(h,h)$
cannot be expressed as a pure boundary \cite{emelson}, it has an
additional bulk part which becomes a constraint on the linear order
perturbation of the metric tensor. The constraint in Einstein's theory
reads 
\begin{equation}
\frac{1}{\Lambda}\intop_{\varSigma}d^{n-1}x\thinspace\sqrt{\bar{\gamma}}\thinspace\bar{\xi}_{\mu}(\Gamma_{\nu\rho}^{\beta})^{\left(1\right)}\bar{\nabla}^{\rho}\bar{\xi}^{\sigma}(\text{\ensuremath{{\cal {P}}}}^{\nu\mu}\thinspace_{\beta\sigma})^{\left(1\right)}=0.
\end{equation}

This paper is organized as follows: in section $\text{\mbox{II}}$
we consider the cosmological Einstein's gravity and give the Abbott-Deser
(AD) formula of the conserved charges \cite{AD} for background Einstein
spacetimes and we summarize the new formulation \cite{newformula,newformulauzun}
to construct the conserved charges. Then we give the linear order
perturbation of the new formula and its behavior under gauge transformations
for (anti) de Sitter background spacetime. In section $\text{\mbox{III}}$,
we discuss the second order perturbations of the new formula and construct
the gauge transformation of the result. In section $\text{\mbox{IV}}$
we discuss the results in terms of second order gauge-invariant perturbation
theory of Nakamura \cite{nakamura2003,nakamura2004,nakamura2007,nakamura2008},
which is a useful technique to construct the relevant quantities as
gauge-variant and invariant parts explicitly. Since the computations
are somewhat lengthy we relegate them to the Appendices.

\section{First order perturbation \i n the cosmological Einstein theory}

The linear order expansion of the cosmological Einstein tensor\footnote{The details of the computations are given at Appendix A.}
about a generic background is

\begin{equation}
(\ensuremath{{\cal {G}}}_{\mu\nu}){}^{\left(1\right)}:=(R_{\mu\nu}){}^{\left(1\right)}-\frac{1}{2}\bar{g}_{\mu\nu}(R){}^{\left(1\right)}-\frac{1}{2}h_{\mu\nu}\bar{R}+\Lambda h_{\mu\nu}.
\end{equation}
This background tensor can be written as two parts \cite{DT,sisman-tekin-setare}
\begin{equation}
\left({\cal G}^{\mu\nu}\right)^{\left(1\right)}=\bar{\nabla}_{\alpha}\bar{\nabla}_{\beta}K^{\mu\alpha\nu\beta}+X^{\mu\nu},
\end{equation}
with
\begin{align}
X^{\mu\nu}\equiv\frac{1}{2}\left(h^{\mu\alpha}\bar{R}_{\alpha}\thinspace^{\nu}-\bar{R}^{\mu\alpha\nu\beta}h_{\alpha\beta}\right)+\frac{1}{2}\bar{g}^{\mu\nu}h^{\rho\sigma}\bar{R}_{\rho\sigma}+\Lambda h^{\mu\nu}-\frac{1}{2}h^{\mu\nu}\bar{R},
\end{align}
and 
\begin{equation}
K^{\mu\alpha\nu\beta}\equiv\frac{1}{2}\left(\bar{g}^{\alpha\nu}\tilde{h}^{\mu\beta}+\bar{g}^{\mu\beta}\tilde{h}^{\alpha\nu}-\bar{g}^{\alpha\beta}\tilde{h}^{\mu\nu}-\bar{g}^{\mu\nu}\tilde{h}^{\alpha\beta}\right).
\end{equation}
Here $\tilde{h}^{\mu\nu}\equiv h^{\mu\nu}-\frac{1}{2}\bar{g}^{\mu\nu}h.$
Let us assume that the background spacetime has at least one Killing
vector field, say $\bar{\xi}_{\nu}$. Contraction of the background
Killing vector $\bar{\xi}_{\nu}$ with $\left({\cal G}^{\mu\nu}\right)^{\left(1\right)}$
yields
\begin{equation}
\bar{\xi}_{\nu}\left({\cal G}^{\mu\nu}\right)^{\left(1\right)}=\bar{\nabla}_{\alpha}\left(\bar{\xi}_{\nu}\bar{\nabla}_{\beta}K^{\mu\alpha\nu\beta}-K^{\mu\beta\nu\alpha}\bar{\nabla}_{\beta}\bar{\xi}_{\nu}\right)+K^{\mu\alpha\nu\beta}\bar{R}_{\thinspace\beta\alpha\nu}^{\rho}\bar{\xi}_{\rho}+X^{\mu\nu}\bar{\xi}_{\nu},\label{onemli}
\end{equation}
where the last two terms vanish for a background \textit{Einstein
spacetime} and, therefore the current can be written as pure divergence
\begin{equation}
\bar{\xi}_{\nu}\left({\cal G}^{\mu\nu}\right)^{\left(1\right)}=\bar{\nabla}_{\mu}\bar{\nabla}_{\alpha}\left(\bar{\xi}_{\nu}\bar{\nabla}_{\beta}K^{\mu\alpha\nu\beta}-K^{\mu\beta\nu\alpha}\bar{\nabla}_{\beta}\bar{\xi}_{\nu}\right):=\bar{\nabla}_{\mu}F^{\mu\nu}.\label{onemli2}
\end{equation}
One natural question is to ask is how this expression changes when
one changes the coordinates on the background spacetime. Under a small
diffeomorphism generated by a vector field $X$, this equation does
not change since $\delta_{X}\left({\cal G}^{\mu\nu}\right)^{\left(1\right)}=\mathscr{L}_{X}{\cal \bar{G}}^{\mu\nu}$,
which vanishes for the background Einstein spaces. Although the result
is gauge-invariant, the antisymmetric tensor $F^{\mu\nu}$ as defined
(\ref{onemli2}) is gauge-invariant only up to a boundary. The change
of $F^{\mu\nu}$ under gauge transformations is complicated and was
given in \cite{newformulauzun}. On the other hand, for (anti) de
Sitter background spacetime it is possible to express the current
in a completely gauge-invariant way \cite{newformula,newformulauzun},
starting from the second Bianchi identity on the Riemann tensor
\begin{equation}
\nabla_{\nu}R_{\sigma\beta\mu\rho}+\nabla_{\sigma}R_{\beta\nu\mu\rho}+\nabla_{\beta}R_{\nu\sigma\mu\rho}=0.
\end{equation}
Using the contracted Bianchi identity $\nabla_{\mu}\text{\ensuremath{{\cal {G}}}}^{\mu\nu}=0$,
the metric compatibility $\nabla_{\mu}g_{\alpha\beta}=0$; and carrying
out the $g^{\nu\rho}$ multiplication, one can construct a divergence-free
rank four tensor (let us denote it as $\text{\ensuremath{{\cal {P}}}}^{\nu\mu}\thinspace_{\beta\sigma}$)
which has additional properties. It has the same symmetries as the
Riemann tensor, it vanishes for the background (anti) de Sitter space,
$\bar{\text{\ensuremath{{\cal {P}}}}}^{\nu\mu}\thinspace_{\beta\sigma}=0$,
its trace is the cosmological Einstein tensor, $\text{\ensuremath{{\cal {P}}}}^{\mu}\thinspace_{\sigma}:=\text{\ensuremath{{\cal {P}}}}^{\nu\mu}\thinspace_{\nu\sigma}=(3-n)\text{\ensuremath{{\cal {G}}}}_{\sigma}^{\mu}$.
Explicitly the $\text{\ensuremath{{\cal {P}}}}$-tensor reads as

\begin{equation}
\text{\ensuremath{{\cal {P}}}}^{\nu\mu}\thinspace_{\beta\sigma}:=R^{\nu\mu}\thinspace_{\beta\sigma}+\delta_{\sigma}^{\nu}\text{\ensuremath{{\cal {G}}}}_{\beta}^{\mu}-\delta_{\beta}^{\nu}\text{\ensuremath{{\cal {G}}}}_{\sigma}^{\mu}+\delta_{\beta}^{\mu}\text{\ensuremath{{\cal {G}}}}_{\sigma}^{\nu}-\delta_{\sigma}^{\mu}\text{\ensuremath{{\cal {G}}}}_{\beta}^{\nu}+\left(\frac{R}{2}-\frac{\Lambda\left(n+1\right)}{n-1}\right)\left(\delta_{\sigma}^{\nu}\delta_{\beta}^{\mu}-\delta_{\beta}^{\nu}\delta_{\sigma}^{\mu}\right).\label{eq:ptensoruudd-1}
\end{equation}
This tensor was used to give a new formulation of conserved charges
in \cite{newformula}, also the construction is improved for the extensions
of the Einstein's gravity in \cite{newformulauzun}. Let us summarize
how one can construct the conserved charges by using the $\text{\ensuremath{{\cal {P}}}}$-tensor.
Consider the following exact equation
\begin{equation}
\nabla_{\nu}(\text{\ensuremath{{\cal {P}}}}^{\nu\mu}\thinspace_{\beta\sigma}\nabla^{\beta}\xi^{\sigma})-\text{\ensuremath{{\cal {P}}}}^{\nu\mu}\thinspace_{\beta\sigma}\nabla_{\nu}\nabla^{\beta}\xi^{\sigma}=0,\label{eq:ddimensionalmainequation-1}
\end{equation}
which is valid for all smooth metrics without the use of the field
equations. Consider the background to be the $n$-dimensional (anti)
de Sitter spacetime with the following equations

\begin{equation}
\bar{R}_{\mu\alpha\nu\beta}=\frac{2}{\left(n-2\right)\left(n-1\right)}\Lambda\left(\bar{g}_{\mu\nu}\bar{g}_{\alpha\beta}-\bar{g}_{\mu\beta}\bar{g}_{\alpha\nu}\right),\qquad\bar{R}_{\mu\nu}=\frac{2}{n-2}\Lambda\bar{g}_{\mu\nu},\qquad\bar{R}=\frac{2n\Lambda}{n-2}.\label{eq:Max_sym_background}
\end{equation}
 First order expansion of (\ref{eq:ddimensionalmainequation-1}) about
the background (anti) de Sitter spacetime gives
\begin{equation}
\bar{\nabla}_{\nu}\biggl((\text{\ensuremath{{\cal {P}}}}^{\nu\mu}\thinspace_{\beta\sigma})^{\left(1\right)}\bar{\nabla}^{\beta}\bar{\xi}^{\sigma}\biggr)-(\text{\ensuremath{{\cal {P}}}}^{\nu\mu}\thinspace_{\beta\sigma})^{\left(1\right)}\bar{\nabla}_{\nu}\bar{\nabla}^{\beta}\bar{\xi}^{\sigma}=0,\label{eq:ddimensionalmainequationlinear}
\end{equation}
where the linear order expansion of the $\text{\ensuremath{{\cal {P}}}}$-tensor
about the (anti) de Sitter spacetime reads 
\begin{equation}
({\cal {P}}^{\nu\mu}\thinspace_{\beta\sigma})^{\left(1\right)}=(R^{\nu\mu}\thinspace_{\beta\sigma})^{(1)}+2(\text{\ensuremath{{\cal {G}}}}_{[\beta}^{\mu})^{(1)}\delta_{\sigma]}^{\nu}+2(\text{\ensuremath{{\cal {G}}}}_{[\sigma}^{\nu})^{(1)}\delta_{\beta]}^{\mu}+(R)^{\left(1\right)}\delta_{[\beta}^{\mu}\delta_{\sigma]}^{\nu}.\label{ktensorlinear}
\end{equation}
Substituting the linearized $\text{\ensuremath{{\cal {P}}}}$-tensor,
assuming $\bar{\xi}^{\mu}$ to be Killing vector and using the identity
$\bar{\nabla}_{\nu}\bar{\nabla}_{\beta}\xi_{\sigma}=\bar{R}_{\lambda\nu\beta\sigma}\bar{\xi}^{\lambda}$,
the linearized equation (\ref{eq:ddimensionalmainequationlinear})
becomes
\begin{equation}
\bar{\text{\ensuremath{\xi}}}_{\nu}(\text{\ensuremath{{\cal {G}}}}^{\nu\mu})^{\left(1\right)}=c\bar{\nabla}_{\nu}\biggl((\text{\ensuremath{{\cal {P}}}}^{\nu\mu}\thinspace_{\beta\sigma})^{\left(1\right)}\bar{\nabla}^{\beta}\bar{\xi}^{\sigma}\biggr),\label{eq:finallinearequation}
\end{equation}
where we have defined $c=\frac{(n-1)(n-2)}{4\Lambda\left(n-3\right)}$.
Since $(\text{\ensuremath{{\cal {G}}}}^{\mu\nu})^{\left(1\right)}$
and $(R)^{(1)}$ vanish on the boundary, the conserved charges of
the cosmological Einstein's theory can be written as
\begin{equation}
Q=\frac{c}{2G\Omega_{n-2}}\int_{\partial\bar{\Sigma}}d^{n-2}x\,\sqrt{\bar{\sigma}}\,\bar{n}_{\mu}\bar{\sigma}_{\nu}\left(R^{\nu\mu}\thinspace_{\beta\sigma}\right)^{\left(1\right)}\bar{\nabla}^{\beta}\bar{\xi}^{\sigma},\label{newcharge}
\end{equation}
where $\bar{\sigma}_{\nu}$ is the unit outward normal vector on the
boundary of the hypersurface, $\partial\bar{\Sigma}$. For a general
background spacetime, under a variation generated by the vector field
$X$ the first order linearized Riemann tensor changes as $\delta_{X}\left(R^{\nu\mu}\thinspace_{\beta\sigma}\right)^{\left(1\right)}=\mathscr{L}_{X}\bar{R}^{\nu\mu}\thinspace_{\beta\sigma}$,
which vanishes for (anti) de Sitter background (for more details see
\cite{newformulauzun}). It turns out, the conserved charges are given
with a gauge-invariant expression which involves the linearized Riemann
tensor explicitly. 

\section{Second order perturbation theory \i n the cosmological Einstein gravity}

Here we discuss the second order perturbations of the cosmological
Einstein tensor following \cite{emelson}. After using the linearized
equation $\bar{\nabla}_{\nu}(\text{\ensuremath{{\cal {P}}}}{}^{\nu\mu}\thinspace_{\beta\sigma})^{\left(1\right)}=0$,
the second order perturbation of equation (\ref{eq:ddimensionalmainequation-1})
about background (anti) de Sitter spacetime reduces to the divergence
and non-divergence parts as
\begin{equation}
\bar{\xi}^{\nu}(\ensuremath{{\cal {G}}}_{\nu}^{\mu})^{\left(2\right)}=c\Biggl(\bar{\nabla}_{\nu}\Bigl(\bar{\nabla}^{\beta}\bar{\xi}^{\sigma}(T^{\nu\mu}\thinspace_{\beta\sigma})^{\left(2\right)}\Bigr)-2(\Gamma_{\nu\rho}^{\beta})^{\left(1\right)}\bar{\nabla}^{\rho}\bar{\xi}^{\sigma}(\text{\ensuremath{{\cal {P}}}}^{\nu\mu}\thinspace_{\beta\sigma})^{\left(1\right)}\Biggr),\label{secondordermainequation}
\end{equation}
where we have defined a second order background tensor
\begin{equation}
(T^{\nu\mu}\thinspace_{\beta\sigma})^{\left(2\right)}:=(\text{\ensuremath{{\cal {P}}}}^{\nu\mu}\thinspace_{\beta\sigma})^{\left(2\right)}+\frac{h}{2}(\text{\ensuremath{{\cal {P}}}}^{\nu\mu}\thinspace_{\beta\sigma})^{\left(1\right)},
\end{equation}
and the constant $c$ was defined below (\ref{eq:finallinearequation}).
Using the explicit form of the cosmological Einstein gravity field
equations, it was shown that the left-hand side of (\ref{secondordermainequation})
cannot be written as a pure divergence term \cite{emelson}. It turns
out, the non-divergence part can involve some divergence terms, but
it cannot be completely written as a divergence term. It is obvious
that, for a manifold $\mathscr{M}$ with a compact hypersurface $\Sigma$
without a boundary, the non-divergence part of (\ref{secondordermainequation})
becomes an integral constraint on the solutions to the first order
linearized equations. Note that if the spacetime $\mathscr{M}$ has
a compact hypersurface \textit{with a boundary}, then we obtain the
equality (\ref{equalityofcharges}), which relates the solutions of
the first order linearized equations to the solutions of the second
order equations. If solutions to the first and the second order perturbed
equations, say $h$ and $k$ respectively, come from linearization
of an exact solution $g$, then the integral constraint is automatically
satisfied for a spacetime manifold $\mathscr{M}$ which has a compact
hypersurface without a boundary. Similarly, if the spacetime $\mathscr{M}$
has a compact hypersurface \textit{with a boundary}, the equality
of the conserved charges (\ref{equalityofcharges}) will also be satisfied.
Otherwise, we say $\bar{g}$ is linearization unstable and the perturbation
theory about it does not make sense.

\section{GAUGE INVARIANT PERTURBATION THEORY}

The second order gauge-invariant perturbation theory was studied in
detail in \cite{nakamura2004,nakamura2007,nakamura2008} and the existence
of the two perturbation parameters are included in \cite{nakamura2003}.
Gauge-invariant perturbation theory is a technique that allows one
to express the tensor fields in terms of gauge-variant and invariant
terms. Of course, one cannot use this method on any arbitrary background
spacetime since the main assumption of the theory is decomposing the
first order metric perturbation as
\begin{equation}
h_{\mu\nu}:=\widetilde{h}_{\mu\nu}+\text{\ensuremath{\mathscr{L}}}_{X}\bar{g}_{\mu\nu},\label{hdenklem}
\end{equation}
here $\widetilde{h}_{\mu\nu}$ denotes the gauge-invariant part, and
the gauge-variant term $\text{\ensuremath{\mathscr{L}}}_{X}\bar{g}_{\mu\nu}$
denotes the Lie derivative of the background metric with respect to
vector field $X$ which is the generator of the gauge transformation.
In the following discussion, we denote the gauge-variant quantities
with a tilde and the background quantities with a bar. If such a decomposition
exists, one can express the linear order perturbation of any tensor
field $T$ as
\begin{equation}
(T)^{\left(1\right)}=(\widetilde{T})^{\left(1\right)}+\text{\ensuremath{\mathscr{L}}}_{X}\bar{T}.
\end{equation}
The second order perturbation of the metric tensor can be expressed
as
\begin{equation}
k_{\mu\nu}:=\frac{1}{2}\widetilde{k}_{\mu\nu}+\text{\ensuremath{\mathscr{L}}}_{X}h_{\mu\nu}+\frac{1}{2}\left(\text{\ensuremath{\mathscr{L}}}_{Y}-\text{\ensuremath{\mathscr{L}}}_{X}^{2}\right)\bar{g}_{\mu\nu},\label{kdenklem}
\end{equation}
where $Y$, just like $X$ generates the gauge transformations. Using
(\ref{hdenklem}, \ref{kdenklem}) the second order perturbation of
any generic tensor field $T$ can be written as
\begin{equation}
(T)^{\left(2\right)}=(\tilde{T})^{\left(2\right)}+\text{\ensuremath{\mathscr{L}}}_{\text{X}}(T)^{\left(1\right)}+\frac{1}{2}\left(\text{\ensuremath{\mathscr{L}}}_{\text{X}}-\text{\ensuremath{\text{\ensuremath{\mathscr{L}}}_{\text{Y}}}}^{2}\right)\overline{T}.
\end{equation}
Note that since the metric tensor involves irreducible gauge-invariant
terms at the first and the second orders, the gauge-invariant part
of any generic tensor field has the same form. Of course, the irreducible
gauge-invariant part of the tensor field only includes $\widetilde{h}_{\mu\nu}$
and $\widetilde{k}_{\mu\nu}$. Details of the calculations are given
in Appendix C. Here we discuss the conserved charges, which are constructed
by using the $\text{\ensuremath{{\cal {P}}}}$-tensor, in terms of
the gauge-invariant perturbation theory. Let us start with the first
order linearized equation (\ref{eq:finallinearequation}), which we
can use to construct the conserved charges. In terms of the gauge-invariant
perturbation theory, the left-hand side of the equation (\ref{eq:finallinearequation})
is gauge-invariant 
\begin{equation}
\bar{\text{\ensuremath{\xi}}}_{\nu}(\text{\ensuremath{{\cal {G}}}}^{\nu\mu})^{\left(1\right)}=\bar{\text{\ensuremath{\xi}}}_{\nu}\left((\tilde{\text{\ensuremath{{\cal {G}}}}}^{\nu\mu})^{\left(1\right)}+\text{\ensuremath{\mathscr{L}}}_{X}\bar{{\cal {G}}}^{\mu\nu}\right)=\bar{\text{\ensuremath{\xi}}}_{\nu}(\tilde{\text{\ensuremath{{\cal {G}}}}}^{\nu\mu})^{\left(1\right)},
\end{equation}
since we consider the (anti) de Sitter background spacetime, for which
we have $\bar{{\cal {G}}}^{\mu\nu}=0$. The right-hand side of (\ref{eq:finallinearequation}),
can be written as
\begin{equation}
\bar{\nabla_{\nu}}\biggl((\text{\ensuremath{{\cal {P}}}}^{\nu\mu}\thinspace_{\beta\sigma})^{\left(1\right)}\bar{\nabla}^{\beta}\bar{\xi}^{\sigma}\biggr)=\bar{\nabla_{\nu}}\biggl(\left((\widetilde{\text{\ensuremath{{\cal {P}}}}}^{\nu\mu}\thinspace_{\beta\sigma})^{\left(1\right)}+\text{\ensuremath{\mathscr{L}}}_{X}\bar{\text{\ensuremath{{\cal {P}}}}}^{\nu\mu}\thinspace_{\beta\sigma}\right)\bar{\nabla}^{\beta}\bar{\xi}^{\sigma}\biggr).
\end{equation}
This reduces to 
\begin{equation}
\bar{\nabla_{\nu}}\biggl((\text{\ensuremath{{\cal {P}}}}^{\nu\mu}\thinspace_{\beta\sigma})^{\left(1\right)}\bar{\nabla}^{\beta}\bar{\xi}^{\sigma}\biggr)=\bar{\nabla_{\nu}}\biggl((\widetilde{\text{\ensuremath{{\cal {P}}}}}^{\nu\mu}\thinspace_{\beta\sigma})^{\left(1\right)}\bar{\nabla}^{\beta}\bar{\xi}^{\sigma}\biggr)
\end{equation}
by using the vanishing of the $\text{\ensuremath{{\cal {P}}}}$-tensor
for the (anti) de Sitter background spacetime, $\bar{\text{\ensuremath{{\cal {P}}}}}^{\nu\mu}\thinspace_{\beta\sigma}=0$.
So, as in the case of the usual perturbation theory the current is
gauge-invariant. At the second order, the left-hand side of the equation
(\ref{secondordermainequation}) is gauge-invariant, since we have
\begin{equation}
(\ensuremath{{\cal {G}}}_{\nu}^{\mu})^{\left(2\right)}=(\widetilde{\ensuremath{{\cal {G}}}}_{\nu}^{\mu})^{\left(2\right)}+\mathscr{L}_{X}(\ensuremath{{\cal {G}}}_{\nu}^{\mu})^{\left(1\right)}+\frac{1}{2}\left(\text{\ensuremath{\mathscr{L}}}_{Y}-\text{\ensuremath{\mathscr{L}}}_{X}^{2}\right)\bar{\ensuremath{{\cal {G}}}}_{\nu}^{\mu},
\end{equation}
which becomes 
\begin{equation}
(\ensuremath{{\cal {G}}}_{\nu}^{\mu})^{\left(2\right)}=(\widetilde{\ensuremath{{\cal {G}}}}_{\nu}^{\mu})^{\left(2\right)},
\end{equation}
where we used $(\ensuremath{{\cal {G}}}_{\nu}^{\mu})^{\left(1\right)}=0=\bar{\ensuremath{{\cal {G}}}}_{\nu}^{\mu}$
in (anti) de Sitter background spacetime. Now let us compute the right-hand
side of (\ref{secondordermainequation}). For the second order perturbation
of the $\text{\ensuremath{{\cal {P}}}}$-tensor, we get

\begin{equation}
(\text{\ensuremath{{\cal {P}}}}^{\nu\mu}\thinspace_{\beta\sigma})^{\left(2\right)}=(\widetilde{\text{\ensuremath{{\cal {P}}}}}^{\nu\mu}\thinspace_{\beta\sigma})^{\left(2\right)}+\text{\ensuremath{\mathscr{L}}}_{X}(\text{\ensuremath{{\cal {P}}}}^{\nu\mu}\thinspace_{\beta\sigma})^{\left(1\right)}+\frac{1}{2}\left(\text{\ensuremath{\mathscr{L}}}_{Y}-\text{\ensuremath{\mathscr{L}}}_{X}^{2}\right)\text{\ensuremath{\bar{{\cal {P}}}}}^{\nu\mu}\thinspace_{\beta\sigma},
\end{equation}
where the last term vanishes at the (anti) de Sitter background spacetime
and so we obtain

\begin{equation}
(\text{\ensuremath{{\cal {P}}}}^{\nu\mu}\thinspace_{\beta\sigma})^{\left(2\right)}=(\widetilde{\text{\ensuremath{{\cal {P}}}}}^{\nu\mu}\thinspace_{\beta\sigma})^{\left(2\right)}+\text{\ensuremath{\mathscr{L}}}_{X}(\widetilde{\text{\ensuremath{{\cal {P}}}}}^{\nu\mu}\thinspace_{\beta\sigma})^{\left(1\right)}.
\end{equation}
Inserting the results in (\ref{secondordermainequation}) we can write

\begin{multline}
\bar{\xi}^{\nu}(\widetilde{\ensuremath{{\cal {G}}}}_{\nu}^{\mu})^{\left(2\right)}=c\bar{\nabla}_{\nu}\left(\bar{\nabla}^{\beta}\bar{\xi}^{\sigma}(\widetilde{\text{\ensuremath{{\cal {P}}}}}^{\nu\mu}\thinspace_{\beta\sigma})^{\left(2\right)}+\bar{\nabla}^{\beta}\bar{\xi}^{\sigma}\text{\ensuremath{\mathscr{L}}}_{X}(\widetilde{\text{\ensuremath{{\cal {P}}}}}^{\nu\mu}\thinspace_{\beta\sigma})^{\left(1\right)}+\frac{h}{2}\bar{\nabla}^{\beta}\bar{\xi}^{\sigma}(\widetilde{\text{\ensuremath{{\cal {P}}}}}^{\nu\mu}\thinspace_{\beta\sigma})^{\left(1\right)}\right)\\
-2c(\Gamma_{\nu\rho}^{\beta})^{\left(1\right)}\bar{\nabla}^{\rho}\bar{\xi}^{\sigma}(\widetilde{\text{\ensuremath{{\cal {P}}}}}^{\nu\mu}\thinspace_{\beta\sigma})^{\left(1\right)},~~~~~~~~~~~~~~~~~~~~~~~~~~~~~~~~~~~~~~~~~~~~~~~~~~~~~~~~~~~~~~~~~~~\label{eq:res.}
\end{multline}
where the left-hand side and the first term on the right-hand side
are already in a gauge-invariant form. Then, let us concentrate on
the gauge-variant terms. The second term reads
\begin{equation}
\bar{\nabla}_{\nu}\left(\bar{\nabla}^{\beta}\bar{\xi}^{\sigma}\text{\ensuremath{\mathscr{L}}}_{X}(\widetilde{\text{\ensuremath{{\cal {P}}}}}^{\nu\mu}\thinspace_{\beta\sigma})^{\left(1\right)}\right)=(\bar{\nabla}_{\nu}\bar{\nabla}^{\beta}\bar{\xi}^{\sigma})\text{\ensuremath{\mathscr{L}}}_{X}(\widetilde{\text{\ensuremath{{\cal {P}}}}}^{\nu\mu}\thinspace_{\beta\sigma})^{\left(1\right)}+\bar{\nabla}^{\beta}\bar{\xi}^{\sigma}\bar{\nabla}_{\nu}\text{\ensuremath{\mathscr{L}}}_{X}(\widetilde{\text{\ensuremath{{\cal {P}}}}}^{\nu\mu}\thinspace_{\beta\sigma})^{\left(1\right)},
\end{equation}
where the first term vanishes after using the identity $\bar{\nabla}_{\nu}\bar{\nabla}^{\beta}\bar{\xi}^{\sigma}=\bar{R}_{\lambda\nu}\thinspace^{\beta\sigma}\bar{\xi}^{\lambda}$,
and then we obtain
\begin{equation}
\bar{\nabla}_{\nu}\left(\bar{\nabla}^{\beta}\bar{\xi}^{\sigma}\text{\ensuremath{\mathscr{L}}}_{X}(\widetilde{\text{\ensuremath{{\cal {P}}}}}^{\nu\mu}\thinspace_{\beta\sigma})^{\left(1\right)}\right)=\bar{\nabla}^{\beta}\bar{\xi}^{\sigma}\bar{\nabla}_{\nu}\text{\ensuremath{\mathscr{L}}}_{X}(\widetilde{\text{\ensuremath{{\cal {P}}}}}^{\nu\mu}\thinspace_{\beta\sigma})^{\left(1\right)}.
\end{equation}
Using the identity (\ref{identitygeneral}) in Appendix B, we get
\begin{align}
 & \bar{\nabla}^{\beta}\bar{\xi}^{\sigma}\bar{\nabla}_{\nu}\text{\ensuremath{\mathscr{L}}}_{X}(\text{\ensuremath{\widetilde{\text{\ensuremath{{\cal {P}}}}}}}^{\nu\mu}\thinspace_{\beta\sigma})^{\left(1\right)}=\bar{\nabla}^{\beta}\bar{\xi}^{\sigma}\Biggl(\text{\ensuremath{\mathscr{L}}}_{X}\bar{\nabla}_{\nu}(\widetilde{\text{\ensuremath{{\cal {P}}}}}^{\nu\mu}\thinspace_{\beta\sigma})^{\left(1\right)}-\delta_{X}(\Gamma_{\nu\lambda}^{\nu})^{\left(1\right)}(\widetilde{\text{\ensuremath{{\cal {P}}}}}^{\lambda\mu}\thinspace_{\beta\sigma})^{\left(1\right)}\nonumber \\
 & ~~~~~~~~~~~~~~~~~~~~~~~~~~~~~~~~~+2\delta_{X}(\Gamma_{\nu\beta}^{\lambda})^{\left(1\right)}(\widetilde{\text{\ensuremath{{\cal {P}}}}}^{\nu\mu}\thinspace_{\lambda\sigma})^{\left(1\right)}\Biggr).\label{eq:sdsefv}
\end{align}
So one has
\begin{equation}
\bar{\nabla}^{\beta}\bar{\xi}^{\sigma}\bar{\nabla}_{\nu}\text{\ensuremath{\mathscr{L}}}_{X}(\text{\ensuremath{\widetilde{\text{\ensuremath{{\cal {P}}}}}}}^{\nu\mu}\thinspace_{\beta\sigma})^{\left(1\right)}=\bar{\nabla}^{\beta}\bar{\xi}^{\sigma}\left(-\delta_{X}(\Gamma_{\nu\lambda}^{\nu})^{\left(1\right)}(\widetilde{\text{\ensuremath{{\cal {P}}}}}^{\lambda\mu}\thinspace_{\beta\sigma})^{\left(1\right)}+2\delta_{X}(\Gamma_{\nu\beta}^{\lambda})^{\left(1\right)}(\widetilde{\text{\ensuremath{{\cal {P}}}}}^{\nu\mu}\thinspace_{\lambda\sigma})^{\left(1\right)}\right),
\end{equation}
where we have used the first order linearization of $\nabla_{\nu}\text{\ensuremath{{\cal {P}}}}^{\nu\mu}\thinspace_{\beta\sigma}=0$
about the (anti) de Sitter background metric. Substituting the results
in (\ref{eq:res.}) and using the decomposition of the linear order
perturbation of the metric tensor (\ref{hdenklem}), we arrive at
\begin{multline}
\bar{\xi}^{\nu}(\widetilde{\ensuremath{{\cal {G}}}}_{\nu}^{\mu})^{\left(2\right)}=c\bar{\nabla}_{\nu}\left(\bar{\nabla}^{\beta}\bar{\xi}^{\sigma}(\widetilde{\text{\ensuremath{{\cal {P}}}}}^{\nu\mu}\thinspace_{\beta\sigma})^{\left(2\right)}+\frac{z}{2}\bar{\nabla}^{\beta}\bar{\xi}^{\sigma}(\widetilde{\text{\ensuremath{{\cal {P}}}}}^{\nu\mu}\thinspace_{\beta\sigma})^{\left(1\right)}+\bar{\nabla}_{\rho}X^{\rho}\bar{\nabla}^{\beta}\bar{\xi}^{\sigma}(\widetilde{\text{\ensuremath{{\cal {P}}}}}^{\nu\mu}\thinspace_{\beta\sigma})^{\left(1\right)}\right)\\
-c\bar{\nabla}^{\beta}\bar{\xi}^{\sigma}\delta_{X}(\Gamma_{\nu\lambda}^{\nu})^{\left(1\right)}(\widetilde{\text{\ensuremath{{\cal {P}}}}}^{\lambda\mu}\thinspace_{\beta\sigma})^{\left(1\right)}+2c(\widetilde{\text{\ensuremath{{\cal {P}}}}}^{\nu\mu}\thinspace_{\lambda\sigma})^{\left(1\right)}\bar{\nabla}^{\beta}\bar{\xi}^{\sigma}\left(\delta_{X}(\Gamma_{\nu\beta}^{\lambda})^{\left(1\right)}-(\Gamma_{\nu\beta}^{\lambda})^{\left(1\right)}\right),\label{eq:res.-1}
\end{multline}
where the last two terms together form a gauge-invariant combination
from the decomposition of the Christoffel connection

\begin{equation}
(\Gamma_{\nu\beta}^{\lambda})^{\left(1\right)}-\delta_{X}(\Gamma_{\nu\beta}^{\lambda})^{\left(1\right)}=(\widetilde{\Gamma}_{\nu\beta}^{\lambda})^{\left(1\right)}.
\end{equation}
Also, after a straightforward calculation one has
\begin{equation}
\bar{\nabla}_{\nu}\left(\bar{\nabla}_{\rho}X^{\rho}\bar{\nabla}^{\beta}\bar{\xi}^{\sigma}(\widetilde{\text{\ensuremath{{\cal {P}}}}}^{\nu\mu}\thinspace_{\beta\sigma})^{\left(1\right)}\right)-\bar{\nabla}^{\beta}\bar{\xi}^{\sigma}\delta_{X}(\Gamma_{\nu\lambda}^{\nu})^{\left(1\right)}(\widetilde{\text{\ensuremath{{\cal {P}}}}}^{\lambda\mu}\thinspace_{\beta\sigma})^{\left(1\right)}=0,
\end{equation}
which proves the vanishing of the gauge-variant terms. Collecting
the pieces together, one ends up with 
\begin{equation}
\bar{\xi}^{\nu}(\widetilde{\ensuremath{{\cal {G}}}}_{\nu}^{\mu})^{\left(2\right)}=c\bar{\nabla}_{\nu}\left(\bar{\nabla}^{\beta}\bar{\xi}^{\sigma}(\widetilde{\text{\ensuremath{{\cal {P}}}}}^{\nu\mu}\thinspace_{\beta\sigma})^{\left(2\right)}+\frac{\widetilde{h}}{2}\bar{\nabla}^{\beta}\bar{\xi}^{\sigma}(\widetilde{\text{\ensuremath{{\cal {P}}}}}^{\nu\mu}\thinspace_{\beta\sigma})^{\left(1\right)}\right)-2c\bar{\nabla}^{\beta}\bar{\xi}^{\sigma}(\widetilde{\text{\ensuremath{{\cal {P}}}}}^{\nu\mu}\thinspace_{\lambda\sigma})^{\left(1\right)}(\widetilde{\Gamma}_{\nu\beta}^{\lambda})^{\left(1\right)},
\end{equation}
where the result involves divergence and non-divergence terms; $\widetilde{h}$
refers to the gauge-variant trace of the metric perturbation. Unlike
the case of usual perturbation theory, the second order cosmological
Einstein tensor is gauge-invariant in this formulation, so are the
conserved charges. For the compact hypersurfaces without a boundary,
vanishing of the last term becomes an integral constraint on solutions
of the first order linearized equations. 

\section{CONCLUSIONS}

The general covariance principle introduces a large gauge degree of
freedom since there is no preferred coordinate system in General Relativity.
In perturbation theory, computing gauge-invariant results plays an
important role since the gauge-variant results can include some unphysical
parts which depend on our choice of the coordinate system. On the
other hand, the second order gauge-invariant perturbation theory allows
a consistent formulation to compute the gauge-invariant parts of the
relevant expressions. In this technique one can construct the relevant
quantities as gauge-variant and invariant parts. So there is no further
need on discuss for the gauge invariance, since the quantities involve
all information that we need. 

In cosmological Einstein's theory, construction of the gauge-invariant
conserved charges is generally done by using the explicit form of
the field equations. The current does not have to be a gauge-invariant
quantity. Of course finding a gauge-invariant current is more valuable
since one only has the physical terms in this case. At the first order,
starting with the second Bianchi identity, one can compute a gauge-invariant
current that involves the Riemann tensor explicitly. At the second
order neither the cosmological Einstein tensor nor the conserved charges
are gauge-invariant. They are only gauge-invariant up to a boundary
term.

In gauge-invariant perturbation theory, at the first order one has
gauge-invariant current and conserved charges as expected. At the
second order, one has a gauge-invariant cosmological Einstein tensor
which is different from the usual perturbation theory case. So, the
conserved charges and the current are all gauge-invariant in this
theory.

\section*{APPENDIX A: SECOND ORDER PERTURBATION THEORY}

Here we give the explicit expressions of the perturbation theory about
the background spacetime $\bar{g}$, up to and including the second
order terms by considering the following metric tensor decomposition 

\begin{equation}
g_{ab}:=\bar{g}_{ab}+\lambda h_{ab}+\lambda^{2}k_{ab},
\end{equation}
where $\lambda$ is a small parameter, $h_{ab}$ and $k_{ab}$ are
the linear and the second order metric tensor perturbations respectively.
Using $g_{ab}g^{bc}=\delta_{a}^{c}$ , we can compute the expansion
of the inverse metric as

\begin{equation}
g^{ab}=\bar{g}^{ab}-\lambda h^{ab}+\lambda^{2}\left(h_{c}^{a}h^{cb}-k^{ab}\right)
\end{equation}
Let $T$ be a generic tensor, it can be perturbed about the background
spacetime $\bar{g}$ as follows

\begin{equation}
T=\bar{T}+\lambda\left(T\right)^{\left(1\right)}+\lambda^{2}\left(T\right)^{\left(2\right)}.
\end{equation}
The Christoffel symbol $\varGamma_{ab}^{c}$ 

\begin{equation}
\varGamma_{ab}^{c}=\frac{1}{2}g^{cd}\Bigl(\partial_{a}g_{bd}+\partial_{b}g_{ad}-\partial_{d}g_{ab}\Bigr),
\end{equation}
is not a tensor quantity but it can be decomposed in the same way

\begin{equation}
\varGamma_{ab}^{c}=\bar{\Gamma}_{ab}^{c}+\lambda(\varGamma_{ab}^{c})^{\left(1\right)}+\lambda^{2}(\varGamma_{ab}^{c})^{(2)}.
\end{equation}
Inserting the given expressions for the metric and its inverse, we
obtain the linear order perturbation of the Christoffel symbol as

\begin{equation}
(\varGamma_{ab}^{c})^{(1)}=\frac{1}{2}\left(\bar{\nabla}_{a}h_{b}^{c}+\bar{\nabla}_{b}h_{a}^{c}-\bar{\nabla}^{c}h_{ab}\right),\label{eq:firstorderconnection}
\end{equation}
and the second order perturbation as

\begin{equation}
(\varGamma_{ab}^{c})^{(2)}=K_{ab}^{c}-h_{d}^{c}(\Gamma_{ab}^{d})^{(1)},\label{eq:secondorderconnetion}
\end{equation}
where we have defined 

\begin{equation}
K_{ab}^{c}=\frac{1}{2}\left(\bar{\nabla}_{a}k_{b}^{c}+\bar{\nabla}_{b}k_{a}^{c}-\bar{\nabla}^{c}k_{ab}\right).\label{secondorderconnectionkpart}
\end{equation}
We can write the linear order perturbation of the Riemann tensor as

\begin{equation}
(R^{a}\thinspace_{bcd}){}^{\left(1\right)}=\bar{\nabla}_{c}(\Gamma_{db}^{a})^{\left(1\right)}-\bar{\nabla}_{d}(\Gamma_{cb}^{a})^{\left(1\right)},\label{eq:firstorderriemann-1}
\end{equation}
and the second order Riemann tensor as

\begin{equation}
(R^{a}\thinspace_{bcd})^{(2)}=\bar{\nabla}_{c}(\Gamma_{bd}^{a})^{(2)}-\bar{\nabla}_{d}(\Gamma_{bc}^{a})^{(2)}+(\Gamma_{bd}^{e}){}^{\left(1\right)}(\Gamma_{ce}^{a}){}^{\left(1\right)}-(\Gamma_{cb}^{e}){}^{\left(1\right)}(\Gamma_{de}^{a}){}^{\left(1\right)},
\end{equation}
which reduces to

\begin{equation}
(R^{a}\thinspace_{bcd})^{(2)}=2\bar{\nabla}_{[c}K_{d]b}^{a}-\bar{\nabla}_{c}\left(h_{e}^{a}(\Gamma_{bd}^{e})^{(1)}\right)+\bar{\nabla}_{d}\left(h_{e}^{a}(\Gamma_{bc}^{a})^{(1)}\right)+(\Gamma_{bd}^{e}){}^{\left(1\right)}(\Gamma_{ce}^{a}){}^{\left(1\right)}-(\Gamma_{cb}^{e}){}^{\left(1\right)}(\Gamma_{de}^{a}){}^{\left(1\right)},\label{eq:secondorderriemann}
\end{equation}
after using the second order Christoffel connection given in (\ref{eq:secondorderconnetion}).
The first and the second order Ricci tensors are obtained from the
contraction, $R_{ab}:=R^{c}\thinspace_{acb}$ , and we get the linear
order perturbation of the Ricci tensor

\begin{equation}
(R_{ab}){}^{\left(1\right)}=\bar{\nabla}_{c}(\Gamma_{ab}^{c}){}^{\left(1\right)}-\bar{\nabla}_{a}(\Gamma_{cb}^{c}){}^{\left(1\right)},
\end{equation}
and the second order Ricci tensor

\begin{equation}
(R_{ab})^{(2)}=2\bar{\nabla}_{[c}K_{a]b}^{c}-\bar{\nabla}_{c}\left(h_{e}^{c}(\Gamma_{ab}^{e})^{(1)}\right)+\bar{\nabla}_{a}\left(h_{e}^{c}(\Gamma_{cb}^{e})^{(1)}\right)+(\Gamma_{ab}^{e}){}^{\left(1\right)}(\Gamma_{ce}^{c}){}^{\left(1\right)}-(\Gamma_{ac}^{e}){}^{\left(1\right)}(\Gamma_{be}^{c}){}^{\left(1\right)}.\label{eq:secondorderriccitensor}
\end{equation}
The first order linearization of the scalar curvature becomes

\begin{equation}
(R)^{\left(1\right)}=\bar{g}^{ab}(R_{ab}){}^{\left(1\right)}-\bar{R}_{ab}h^{ab},\label{eq:firstorderscalarcurvature}
\end{equation}
and the second order Ricci scalar is

\begin{equation}
(R)^{(2)}=\bar{R}_{ab}\left(h_{c}^{a}h^{bc}-k^{ab}\right)-(R_{ab}){}^{\left(1\right)}h^{ab}+\bar{g}^{ab}(R_{ab})^{(2)}.\label{eq:secondorderscalarcurvature}
\end{equation}
The cosmological Einstein tensor

\begin{equation}
{\cal {G}}_{ab}=R_{ab}-\frac{1}{2}g_{ab}R+\varLambda g_{ab},
\end{equation}
at first order yields
\begin{equation}
({\cal {G}}_{ab}){}^{\left(1\right)}=(R_{ab}){}^{\left(1\right)}-\frac{1}{2}\bar{g}_{ab}(R){}^{\left(1\right)}-\frac{1}{2}\bar{R}h_{ab}+\Lambda h_{ab},\label{firstordercosmologicaleinstein}
\end{equation}
and at the second order becomes

\begin{equation}
({\cal {G}}_{ab}){}^{\left(2\right)}=(R_{ab}){}^{\left(2\right)}-\frac{1}{2}\left(\bar{g}_{ab}(R){}^{\left(2\right)}+h_{ab}(R){}^{\left(1\right)}+k_{ab}\bar{R}+2\Lambda k_{ab}\right).\label{eq:secondordercosmologicaleinstein}
\end{equation}

\section*{APPENDIX B: IDENTITIES ON LIE AND COVARIANT DERIVATIVES}

Lie derivative plays an important role in the second order gauge-invariant
perturbation theory and also in the usual gauge transformations generated
by a vector field. Here we derive some useful identities which heavily
used in the computations. Since Lie and covariant derivatives do not
commute, we need to introduce the expressions in a compact way, that
appears when we change the order of these differentiations. In order
to obtain the desired expressions, let us start with Lie derivative
of a rank two tensor $T$

\begin{equation}
\text{\ensuremath{\mathscr{L}}}_{X}T_{ab}=X^{f}\bar{\nabla}_{f}T_{ab}+T_{fb}\bar{\nabla}_{a}X^{f}+T_{fa}\bar{\nabla}_{b}X^{f}.
\end{equation}
Covariant derivative of this expression yields
\begin{align}
\bar{\nabla}_{c}\text{\ensuremath{\mathscr{L}}}_{X}T_{ab}=\bar{\nabla}_{c}X^{f}\bar{\nabla}_{f}T_{ab}+X^{f}\bar{\nabla}_{c}\bar{\nabla}_{f}T_{ab}+T_{fb}\bar{\nabla}_{c}\bar{\nabla}_{a}X^{f}+\bar{\nabla}_{a}X^{f}\bar{\nabla}_{c}T_{fb}\nonumber \\
+\bar{\nabla}_{c}\bar{\nabla}_{b}X^{f}T_{fa}+\bar{\nabla}_{b}X^{f}\bar{\nabla}_{c}T_{fa}.~~~~~~~~~~~~~~~~~~~~~~~~~~~~~~~~~~~~~\label{appendixb2}
\end{align}
When we change the order of the derivatives we get
\begin{equation}
\text{\ensuremath{\mathscr{L}}}_{X}\bar{\nabla}_{c}T_{ab}=X^{f}\bar{\nabla}_{f}\bar{\nabla}_{c}T_{ab}+\left(\bar{\nabla}_{c}X^{f}\right)\bar{\nabla}_{f}T_{ab}+\left(\bar{\nabla}_{a}X^{f}\right)\bar{\nabla}_{c}T_{fb}+\left(\bar{\nabla}_{b}X^{f}\right)\bar{\nabla}_{c}T_{af},
\end{equation}
and subtraction of the results yields

\begin{equation}
\bar{\nabla}_{c}\text{\ensuremath{\mathscr{L}}}_{X}T_{ab}-\text{\ensuremath{\mathscr{L}}}_{X}\bar{\nabla}_{c}T_{ab}=X^{f}\left[\bar{\nabla}_{c},\bar{\nabla}_{f}\right]T_{ab}+\left(\bar{\nabla}_{c}\bar{\nabla}_{a}X^{f}\right)T_{fb}+\left(\bar{\nabla}_{c}\bar{\nabla}_{b}X^{f}\right)T_{af}.\label{appendixb4}
\end{equation}
Using

\begin{equation}
\left[\bar{\nabla}_{c},\bar{\nabla}_{f}\right]T_{ab}=\overline{R}_{cfa}\thinspace^{e}T_{eb}+\overline{R}_{cfb}\thinspace^{e}T_{ae},
\end{equation}
one can rewrite (\ref{appendixb4}) as

\begin{equation}
\bar{\nabla}_{c}\text{\ensuremath{\mathscr{L}}}_{X}T_{ab}=\text{\ensuremath{\mathscr{L}}}_{X}\bar{\nabla}_{c}T_{ab}+\left(\bar{\nabla}_{c}\bar{\nabla}_{a}X^{e}+\overline{R}_{cfa}\thinspace^{e}X^{f}\right)T_{eb}+\left(\overline{R}_{cfb}\thinspace^{e}X^{f}+\bar{\nabla}_{c}\bar{\nabla}_{b}X^{e}\right)T_{ae}.\label{appendixb6}
\end{equation}
We can relate the last expression with the gauge transformation of
the linearized Christoffel connection as follows. Recall that under
the gauge transformations generated by the vector field $X$, the
linear order metric perturbation transforms as $\delta_{X}h_{ab}=\bar{\nabla}_{a}X_{b}+\bar{\nabla}_{b}X_{a}=\text{\ensuremath{\mathscr{L}}}_{X}\bar{g}_{ab}$,
then the gauge transformation of the linearized Christoffel symbol
becomes

\begin{equation}
\delta_{X}(\Gamma_{ab}^{c}){}^{\left(1\right)}=\frac{1}{2}\left(\bar{\nabla}_{a}\delta_{X}h_{b}^{c}+\bar{\nabla}_{b}\delta_{X}h_{a}^{c}-\bar{\nabla}^{c}\delta_{X}h_{ab}\right),
\end{equation}
which can be rewritten as
\begin{equation}
\delta_{X}(\Gamma_{ab}^{c})^{\left(1\right)}=\bar{\nabla}_{a}\bar{\nabla}_{b}X^{c}+\bar{R}^{c}\thinspace_{bda}X^{d}.\label{gaugetranschristoffel}
\end{equation}
Using the last expression, (\ref{appendixb6}) can be expressed as

\begin{equation}
\bar{\nabla}_{c}\text{\ensuremath{\mathscr{L}}}_{X}T_{ab}=\text{\ensuremath{\mathscr{L}}}_{X}\bar{\nabla}_{c}T_{ab}+\delta_{X}(\Gamma_{ca}^{e})^{\left(1\right)}T_{eb}+\delta_{X}(\Gamma_{cb}^{e})^{\left(1\right)}T_{ae}.\label{eq:identity1}
\end{equation}
Similar computation for a $(1,1)$ tensor ends up with

\begin{equation}
\bar{\nabla}_{c}\text{\ensuremath{\mathscr{L}}}_{X}T^{a}\thinspace_{b}=\text{\ensuremath{\mathscr{L}}}_{X}\bar{\nabla}_{c}T^{a}\thinspace_{b}+T^{a}\thinspace_{e}\delta_{X}(\Gamma_{cb}^{e})^{\left(1\right)}-T^{e}\thinspace_{b}\delta_{X}(\Gamma_{ce}^{a})^{\left(1\right)}.
\end{equation}
We can extend the computation for a general $(m,n)$ tensor as
\begin{align}
 & \bar{\nabla}_{c}\text{\ensuremath{\mathscr{L}}}_{X}T^{a_{1}a_{2}...a_{m}}\thinspace_{b_{1}b_{2}...b_{n}}=\text{\ensuremath{\mathscr{L}}}_{X}\bar{\nabla}_{c}T^{a_{1}a_{2}...a_{m}}\thinspace_{b_{1}b_{2}...b_{n}}~~~~~~~~~~~~~~~~~~~~~~~~~~~~~~~~~~~~~~~~~~~~~~~~~~~~\label{identitygeneral}\\
 & +\delta_{X}(\Gamma_{cb_{1}}^{d})^{\left(1\right)}T^{a_{1}a_{2}...a_{m}}\thinspace_{db_{2}...b_{n}}+\delta_{X}(\Gamma_{cb_{2}}^{d})^{\left(1\right)}T^{a_{1}a_{2}...a_{m}}\thinspace_{b_{1}d...b_{n}}+...+\delta_{X}(\Gamma_{cb_{n}}^{d})^{\left(1\right)}T^{a_{1}a_{2}...a_{m}}\thinspace_{b_{1}b_{2}...d}\nonumber \\
 & -\delta_{X}(\Gamma_{cd}^{a_{1}})^{\left(1\right)}T^{da_{2}...a_{m}}\thinspace_{b_{1}b_{2}...b_{n}}-\delta_{X}(\Gamma_{cd}^{a_{2}})^{\left(1\right)}T^{a_{1}d...a_{m}}\thinspace_{b_{1}b_{2}...b_{n}}-...-\delta_{X}(\Gamma_{cd}^{a_{m}})^{\left(1\right)}T^{a_{1}a_{2}...d}\thinspace_{b_{1}b_{2}...b_{n}},\nonumber 
\end{align}
which simplifies the computations.

\section*{APPENDIX C: SECOND ORDER GAUGE INVARIANT PERTURBATION THEORY}

Here we summarize the results of the second order gauge-invariant
perturbation theory following \cite{nakamura2007}. The gauge transformation
of a physical quantity $T$ reads

\begin{equation}
T(p)=\bar{T}(\bar{p})+\delta T(p)
\end{equation}
here $T(p)$ denotes the physical quantity on spacetime $\mathscr{M}$
at point $p$, $\bar{T}(\bar{p})$ denotes the same quantity on the
background spacetime $\mathscr{M}_{0}$ at point $\bar{p}$ and $\delta T(p)$
denotes the deviation of $T(p)$ from its background value $\bar{T}(\bar{p})$.
We show the metric on $\mathscr{M}$ with $g$ and the metric on the
background spacetime $\mathscr{M}_{0}$ with $\bar{g}$. Let $X$
and $Y$ denote two different gauge choices and let $\xi_{1}$ and
$\xi_{2}$ denote the generators of the gauge transformations. One
can compute the following difference

\begin{equation}
(T)_{Y}^{\left(1\right)}-(T)_{X}^{\left(1\right)}=\text{\ensuremath{\mathscr{L}}}_{\text{\ensuremath{\xi}}_{1}}\overline{T},\label{transformation1}
\end{equation}
where $(T)_{Y}^{\left(1\right)}$ is the linear order perturbation
of the physical quantity $T(p)$ in the gauge $Y$ and $(T)_{X}^{\left(1\right)}$
denotes the same quantity in the gauge $X$. For the second order
perturbation of the physical quantity $T(p)$ we have a similar expression 

\begin{equation}
(T)_{Y}^{(2)}-(T)_{X}^{(2)}=\text{\ensuremath{\mathscr{L}}}_{\xi_{1}}(T)_{X}^{\left(1\right)}+\left(\text{\ensuremath{\mathscr{L}}}_{\xi_{2}}+\text{\ensuremath{\mathscr{L}}}_{\xi_{1}}^{2}\right)\overline{T},
\end{equation}
which shows the difference of the perturbations under the change of
the coordinate system. The generators $\xi_{1}$ and $\xi_{2}$ can
be expressed as follows
\begin{equation}
\xi_{1}:=Y-X
\end{equation}
and
\begin{equation}
\xi_{2}:=\left[Y,X\right],
\end{equation}
note that $\xi_{1}$ and $\xi_{2}$ may be different. Following Nakamura
\cite{nakamura2007}, we assume that the linear order metric perturbation
can be decomposed to gauge-variant and invariant parts as

\begin{equation}
h_{ab}:=\widetilde{h}_{ab}+\bar{\nabla}_{a}X_{b}+\bar{\nabla}_{b}X_{a}=\widetilde{h}_{ab}+\text{\ensuremath{\mathscr{L}}}_{X}\bar{g}_{ab},\label{linearordermetrcidecomposition}
\end{equation}
where $\widetilde{h}_{ab}$ is gauge-invariant term and the $\text{\ensuremath{\mathscr{L}}}_{X}\bar{g}_{ab}$
denotes the gauge-variant part. From the gauge transformation (\ref{transformation1}),
we can write

\begin{equation}
\delta_{Y}\widetilde{h}_{ab}-\delta_{X}\widetilde{h}_{ab}=0,
\end{equation}
which shows the gauge invariance of the $\widetilde{h}_{ab}$. Note
that this assumption depends on the properties of the background spacetime.
If we accept this decomposition, the second order metric perturbation
can be expressed as

\begin{equation}
2k_{ab}:=\widetilde{k}_{ab}+2\text{\ensuremath{\mathscr{L}}}_{X}h_{ab}+\left(\text{\ensuremath{\mathscr{L}}}_{Y}-\text{\ensuremath{\mathscr{L}}}_{X}^{2}\right)\bar{g}_{ab},\label{eq:secondordermetric}
\end{equation}
where $\widetilde{k}_{ab}$ is the gauge-invariant part and the additional
terms are all gauge-variant. Using the given decompositions of the
first and the second order metric perturbations, the linear order
perturbation of a generic tensor field reads

\begin{equation}
(T)^{\left(1\right)}=(\widetilde{T})^{(1)}+\text{\ensuremath{\mathscr{L}}}_{\text{X}}\overline{T},
\end{equation}
which means gauge-variant part of the tensor field is equivalent to
the Lie derivative of this tensor field evaluated at the background
spacetime. For the second order perturbations, we obtain a similar
expression as

\begin{equation}
(T)^{\left(2\right)}=(\tilde{T})^{\left(2\right)}+\text{\ensuremath{\mathscr{L}}}_{\text{X}}(T)^{\left(1\right)}+\frac{1}{2}\left(\text{\ensuremath{\mathscr{L}}}_{\text{X}}-\text{\ensuremath{\text{\ensuremath{\mathscr{L}}}_{\text{Y}}}}^{2}\right)\overline{T}.
\end{equation}
Here $(\tilde{T})^{\left(2\right)}$ is the gauge-variant part of
the second order tensor $(T)^{\left(2\right)}$ and the remaining
terms are gauge-variant. Using (\ref{linearordermetrcidecomposition}),
the linear order perturbation of the Christoffel symbol (\ref{eq:firstorderconnection}),
can be written as
\begin{align}
(\Gamma_{ab}^{c}){}^{\left(1\right)}=\frac{1}{2}\Biggl(\bar{\nabla}_{a}(\widetilde{h}_{b}^{c}+\bar{\nabla}_{b}X^{c}+\bar{\nabla}^{c}X_{b})+\bar{\nabla}_{b}(\widetilde{h}_{a}^{c}+\bar{\nabla}_{a}X^{c}+\bar{\nabla}^{c}X_{a})\nonumber \\
-\bar{\nabla}^{c}(\widetilde{h}_{ab}+\bar{\nabla}_{a}X_{b}+\bar{\nabla}_{b}X_{a})\Biggr).~~~~~~~~~~~~~~~~~~~~~~~~~~~~~~~~~~~~\label{abc}
\end{align}
For simplicity, let us define a new gauge-invariant background tensor

\begin{equation}
(\widetilde{\Gamma}_{ab}^{c})^{\left(1\right)}=\frac{1}{2}\left(\bar{\nabla}_{a}\widetilde{h}_{b}^{c}+\bar{\nabla}_{b}\widetilde{h}_{a}^{c}-\bar{\nabla}^{c}\widetilde{h}_{ab}\right).
\end{equation}
Then we have

\begin{equation}
(\Gamma_{ab}^{c}){}^{\left(1\right)}=(\widetilde{\Gamma}_{ab}^{c})^{\left(1\right)}+\frac{1}{2}\left(2\bar{\nabla}_{a}\bar{\nabla}_{b}X^{c}+\left[\bar{\nabla}_{a},\bar{\nabla}^{c}\right]X_{b}+\left[\bar{\nabla}_{b},\bar{\nabla}_{a}\right]X^{c}+\left[\bar{\nabla}_{b},\bar{\nabla}^{c}\right]X_{a}\right),
\end{equation}
which reduces to 
\begin{equation}
(\Gamma_{ab}^{c}){}^{\left(1\right)}=(\widetilde{\Gamma}_{ab}^{c}){}^{\left(1\right)}+\bar{\nabla}_{a}\bar{\nabla}_{b}X^{c}+\bar{R}^{c}\thinspace_{bda}X^{d},
\end{equation}
where we used the identity $\left[\bar{\nabla}_{a},\bar{\nabla}_{b}\right]X^{c}=\bar{R}_{ab}\thinspace^{cd}X_{d}$,
and the first Bianchi identity $\bar{R}_{abcd}+\bar{R}_{bcad}+\bar{R}_{cabd}=0$.
Furthermore, from (\ref{gaugetranschristoffel}) we get

\begin{equation}
(\Gamma_{ab}\thinspace^{c}){}^{\left(1\right)}=(\widetilde{\Gamma}_{ab}\thinspace^{c}){}^{\left(1\right)}+\delta_{X}(\Gamma_{ab}\thinspace^{c}){}^{\left(1\right)},
\end{equation}
which relates the linearized Christoffel connection with the usual
gauge transformation of the linearized Christoffel symbol generated
by the vector field $X$. Similarly the first order expansion of the
Riemann tensor (\ref{eq:firstorderriemann-1}) can be expressed as

\begin{equation}
(R^{a}\thinspace_{bcd}){}^{\left(1\right)}=2\bar{\nabla}_{[c}(\widetilde{\Gamma}_{d]b}^{a})^{\left(1\right)}+\left[\bar{\nabla}_{c},\bar{\nabla}_{d}\right]\bar{\nabla}_{b}X^{a}+\bar{R}^{a}\thinspace_{bed}\bar{\nabla}_{c}X^{e}-\bar{R}^{a}\thinspace_{bec}\bar{\nabla}_{d}X^{e}+X^{e}(\bar{\nabla}_{c}\bar{R}^{a}\thinspace_{bed}-\bar{\nabla}_{d}\bar{R}^{a}\thinspace_{bec})
\end{equation}
and reduces to
\begin{equation}
(R^{a}\thinspace_{bcd}){}^{\left(1\right)}=2\bar{\nabla}_{[c}(\widetilde{\Gamma}_{d]b}^{a})^{\left(1\right)}+X^{e}\bar{\nabla}_{e}\bar{R}^{a}\thinspace_{bcd}+\bar{R}^{a}\thinspace_{bed}\bar{\nabla}_{c}X^{e}+\bar{R}^{a}\thinspace_{bce}\bar{\nabla}_{d}X^{e}+\bar{R}^{a}\thinspace_{ecd}\bar{\nabla}_{b}X^{e}-\bar{R}^{e}\thinspace_{bcd}\bar{\nabla}_{e}X^{a},
\end{equation}
after using the second Bianchi identity $\bar{\nabla}_{a}\bar{R}_{bcde}+\bar{\nabla}_{b}\bar{R}_{cade}+\bar{\nabla}_{c}\bar{R}_{abde}=0$.
Note that the gauge-variant part is obviously given as the Lie derivative
of the Riemann tensor evaluated at the background spacetime. Then
the final expression becomes

\begin{equation}
(R^{a}\thinspace_{bcd}){}^{\left(1\right)}=2\bar{\nabla}_{[c}(\widetilde{\Gamma}_{d]b}^{a})^{\left(1\right)}+\text{\ensuremath{\mathscr{L}}}_{X}\bar{R}^{a}\thinspace_{bcd},\label{eq:firstordergaugeinvriemann}
\end{equation}
which is consistent with the aim of the gauge-invariant perturbation
theory. The first order linearized Ricci tensor can be found from
the contraction of the first and the third indices, $(R_{ab}){}^{\left(1\right)}:=(R^{c}\thinspace_{acb}){}^{\left(1\right)}$,
so we have

\begin{equation}
(R_{ab}){}^{\left(1\right)}=2\bar{\nabla}_{[c}(\widetilde{\Gamma}_{a]b}^{c})^{\left(1\right)}+\text{\ensuremath{\mathscr{L}}}_{X}\bar{R}_{ab}.
\end{equation}
 Since the first order linearized Christoffel connection is a background
tensor, we can lower and raise the indices with the background metric
and the inverse metric respectively. For an example we use $(\Gamma_{acd})^{\left(1\right)}:=\bar{g}_{bd}(\Gamma_{ac}^{b})^{\left(1\right)}$,
where the up index is lowered as the last down index. The first order
linearized scalar curvature, by using (\ref{eq:firstorderscalarcurvature})
and the previous results, becomes

\begin{equation}
(R)^{\left(1\right)}=2\bar{\nabla}_{[b}(\widetilde{\Gamma}_{a]}\thinspace^{ab})^{\left(1\right)}+\bar{g}^{ab}\text{\ensuremath{\mathscr{L}}}_{X}\bar{R}_{ab}-\bar{R}_{ab}(\widetilde{h}^{ab}-\text{\ensuremath{\mathscr{L}}}_{X}\bar{g}^{ab}).
\end{equation}
Equivalently, it can be written as

\begin{equation}
(R)^{\left(1\right)}=2\bar{\nabla}_{[b}(\widetilde{\Gamma}_{a]}\thinspace^{ab})^{\left(1\right)}-\bar{R}_{ab}\widetilde{h}^{ab}+\text{\ensuremath{\mathscr{L}}}_{X}(\bar{R}).
\end{equation}
Inserting the corresponding expressions in the first order linearized
cosmological Einstein tensor (\ref{firstordercosmologicaleinstein}),
we get

\begin{equation}
({\cal {G}}_{ab}){}^{\left(1\right)}=2\bar{\nabla}_{[c}(\widetilde{\Gamma}_{a]b}\thinspace^{c})^{\left(1\right)}+\bar{g}_{ab}\bar{\nabla}_{[c}(\widetilde{\Gamma}_{d]}{}^{cd})^{\left(1\right)}+\frac{1}{2}\bar{g}_{ab}\bar{R}_{cd}\widetilde{h}^{cd}+\widetilde{h}_{ab}\left(\Lambda-\frac{1}{2}\bar{R}\right)+\text{\ensuremath{\mathscr{L}}}_{X}\bar{{\cal {G}}}_{ab},
\end{equation}
where only the last term is gauge-variant and it vanishes if $\bar{g}$
is a background solution, if this is the case $({\cal {G}}_{ab}){}^{\left(1\right)}$
becomes gauge-invariant.

Now, we compute the decompositions of the second order tensors in
terms of gauge-variant and invariant parts. We can compute (\ref{secondorderconnectionkpart})
by using (\ref{eq:secondordermetric}) as

\begin{align}
K_{ab}^{c}=\frac{1}{4}(\bar{\nabla}_{a}\widetilde{k}_{b}^{c}+\bar{\nabla}_{b}\widetilde{k}_{a}^{c}-\bar{\nabla}^{c}\widetilde{k}_{ab})~~~~~~~~~~~~~~~~~~~~~~~~~~~~~~~~~~~~~~~~~~~~~~~~~~~~~~~~~~~~~\label{threeindexkexpression}\\
~~~~~~~~~+\frac{1}{4}\overline{g}^{cd}\left(\bar{\nabla}_{a}\text{\ensuremath{\mathscr{L}}}_{X}\left(h_{bd}+\widetilde{h}_{bd}\right)+\bar{\nabla}_{b}\text{\ensuremath{\mathscr{L}}}_{X}\left(h_{ad}+\widetilde{h}_{ad}\right)-\bar{\nabla}_{d}\text{\ensuremath{\mathscr{L}}}_{X}\left(h_{ab}+\widetilde{h}_{ab}\right)\right)\nonumber \\
+\frac{1}{4}\overline{g}^{cd}\left(\bar{\nabla}_{a}\text{\ensuremath{\mathscr{L}}}_{Y}\overline{g}_{bd}+\bar{\nabla}_{b}\text{\ensuremath{\mathscr{L}}}_{Y}\overline{g}_{ad}-\bar{\nabla}_{d}\text{\ensuremath{\mathscr{L}}}_{Y}\overline{g}_{ab}\right).~~~~~~~~~~~~~~~~~~~~~~~~~~~~~~~~~~~~\nonumber 
\end{align}
After defining a new gauge-invariant second order background tensor

\begin{equation}
\widetilde{K}_{ab}^{c}=\frac{1}{2}\left(\bar{\nabla}_{a}\widetilde{k}_{b}^{c}+\bar{\nabla}_{b}\widetilde{k}_{a}^{c}-\bar{\nabla}^{c}\widetilde{k}_{ab}\right),
\end{equation}
we obtain
\begin{align}
2K_{ab}^{c}=\widetilde{K}_{ab}^{c}+\frac{1}{2}\overline{g}^{cd}\text{\ensuremath{\mathscr{L}}}_{X}\left(\bar{\nabla}_{a}\left(h_{bd}+\widetilde{h}_{bd}\right)+\bar{\nabla}_{b}\left(h_{ad}+\widetilde{h}_{ad}\right)-\bar{\nabla}_{d}\left(h_{ab}+\widetilde{h}_{ab}\right)\right)\nonumber \\
+\left(h_{e}^{c}+\widetilde{h}_{e}^{c}\right)\delta_{X}\left(\Gamma_{ab}^{e}\right)^{\left(1\right)}+\delta_{Y}\left(\Gamma_{ab}^{c}\right)^{\left(1\right)}.~~~~~~~~~~~~~~~~~~~~~~~~~~~~~~~~~~~~~~~~~~~\label{threeindexk2}
\end{align}
Note that we have used the identity (\ref{eq:identity1}) given in
Appendix B to get the last expression. After a straightforward calculation
the result reduces to

\begin{align}
2K_{ab}^{c}=\widetilde{K}_{ab}^{c}+\text{\ensuremath{\mathscr{L}}}_{X}\left(\left(\Gamma_{ab}^{c}\right)^{\left(1\right)}+\left(\widetilde{\Gamma}_{ab}^{c}\right)^{\left(1\right)}\right)-\text{\ensuremath{\mathscr{L}}}_{X}\overline{g}^{cd}\left(\left(\Gamma_{abd}\right)^{\left(1\right)}+\left(\widetilde{\Gamma}_{abd}\right)^{\left(1\right)}\right)\nonumber \\
+\left(h_{e}^{c}+\widetilde{h}_{e}^{c}\right)\delta_{X}\left(\Gamma_{ab}^{e}\right)^{\left(1\right)}+\delta_{Y}\left(\Gamma_{ab}^{e}\right)^{\left(1\right)}.~~~~~~~~~~~~~~~~~~~~~~~~~~~~~~~~~~~~~~~~\label{threeindexkfinal}
\end{align}
We can construct the following tensor 

\begin{multline}
\bar{4\nabla}_{[c}K_{d]b}^{a}=\bar{2\nabla}_{[c}\widetilde{K}_{d]b}^{a}+\bar{\nabla}_{c}\left(\text{\ensuremath{\mathscr{L}}}_{X}\bigl((\Gamma_{bd}^{a}){}^{\left(1\right)}+(\widetilde{\Gamma}_{bd}^{a}){}^{\left(1\right)}\bigr)\right)-\bar{\nabla}_{c}\left(\text{\ensuremath{\mathscr{L}}}_{X}\overline{g}^{ea}\bigl((\Gamma_{bde})^{\left(1\right)}+(\widetilde{\Gamma}_{bde})^{\left(1\right)}\bigr)\right)\\
+\bar{\nabla}_{c}\left(\left(h_{e}^{a}+\widetilde{h}_{e}^{a}\right)\delta_{X}\left(\Gamma_{bd}^{e}\right)^{\left(1\right)}\right)+\bar{\nabla}_{c}\delta_{Y}\left(\Gamma_{bd}^{e}\right)^{\left(1\right)}~~~~~~~~~~~~~~~~~~~~~~~~~~~~~~~~~~~~\\
-\bar{\nabla}_{d}\left(\text{\ensuremath{\mathscr{L}}}_{X}\left((\Gamma_{bc}^{a}){}^{\left(1\right)}+(\widetilde{\Gamma}_{bc}^{a}){}^{\left(1\right)}\right)\right)+\bar{\nabla}_{d}\left(\text{\ensuremath{\mathscr{L}}}_{X}\overline{g}^{ea}\bigl((\Gamma_{bce})^{\left(1\right)}+(\widetilde{\Gamma}_{bce})^{\left(1\right)}\bigr)\right)~~~~~\\
-\bar{\nabla}_{d}\left(\left(h_{e}^{a}+\widetilde{h}_{e}^{a}\right)\delta_{X}\left(\Gamma_{bc}^{e}\right)^{\left(1\right)}\right)-\bar{\nabla}_{d}\delta_{Y}\left(\Gamma_{bc}^{a}\right)^{\left(1\right)},~~~~~~~~~~~~~~~~~~~~~~~~~~~~~~~~~~~~~~~~~~~~~\label{covariantderivativethreeindexk}
\end{multline}
to compute the the second order perturbation of the Riemann tensor
(\ref{eq:secondorderriemann}). Using (\ref{identitygeneral}), it
can be written as

\begin{multline}
\bar{4\nabla}_{[c}K_{d]b}^{a}=\bar{2\nabla}_{[c}\widetilde{K}_{d]b}^{a}+\text{\ensuremath{\mathscr{L}}}_{X}\left(\bar{\nabla}_{c}\bigl((\Gamma_{bd}^{a}){}^{\left(1\right)}+(\widetilde{\Gamma}_{bd}^{a}){}^{\left(1\right)}\bigr)-\bar{\nabla}_{d}\bigl((\Gamma_{bc}^{a}){}^{\left(1\right)}+(\widetilde{\Gamma}_{bc}^{a}){}^{\left(1\right)}\bigr)\right)\\
+(h_{e}^{a}+\widetilde{h}_{e}^{a})\left(\bar{\nabla}_{c}\delta_{X}(\Gamma_{bd}^{e})^{\left(1\right)}-\bar{\nabla}_{d}\delta_{X}(\Gamma_{bc}^{e})^{\left(1\right)}\right)+\bar{\nabla}_{c}\delta_{Y}(\Gamma_{bd}^{a})^{\left(1\right)}-\bar{\nabla}_{d}\delta_{Y}(\Gamma_{bc}^{a})^{\left(1\right)}\\
+\left((\Gamma_{ed}^{a}){}^{\left(1\right)}+(\widetilde{\Gamma}_{ed}^{a}){}^{\left(1\right)}-\bar{\nabla}_{d}\bigl(h_{e}^{a}+\widetilde{h}_{e}^{a}\bigr)\right)\delta_{X}(\Gamma_{cb}^{e}){}^{\left(1\right)}~~~~~~~~~~~~~~~~~~~~~~~~~~~~~~~~~~\\
-\left((\Gamma_{ec}^{a}){}^{\left(1\right)}+(\widetilde{\Gamma}_{ec}^{a}){}^{\left(1\right)}-\bar{\nabla}_{c}\bigl(h_{e}^{a}+\widetilde{h}_{e}^{a}\bigr)\right)\delta_{X}(\Gamma_{db}^{e}){}^{\left(1\right)}~~~~~~~~~~~~~~~~~~~~~~~~~~~~~~~~~~\\
-\left((\Gamma_{bd}^{e}){}^{\left(1\right)}+(\widetilde{\Gamma}_{bd}^{e}){}^{\left(1\right)}\right)\left(\delta_{X}(\Gamma_{ce}^{a}){}^{\left(1\right)}-\bar{\nabla}_{c}\bigl(\bar{\nabla}^{a}X_{e}+\bar{\nabla}_{e}X^{a}\bigr)\right)~~~~~~~~~~~~~~~~~~~~~\\
+\left((\Gamma_{bc}^{e}){}^{\left(1\right)}+(\widetilde{\Gamma}_{bc}^{e}){}^{\left(1\right)}\right)\left(\delta_{X}(\Gamma_{de}^{a}){}^{\left(1\right)}-\bar{\nabla}_{d}\bigl(\bar{\nabla}^{a}X_{e}+\bar{\nabla}_{e}X^{a}\bigr)\right)~~~~~~~~~~~~~~~~~~~~~\\
-\text{\ensuremath{\mathscr{L}}}_{X}\overline{g}^{ea}\left(\bar{\nabla}_{c}\bigl((\Gamma_{bde})^{\left(1\right)}+(\widetilde{\Gamma}_{bde})^{\left(1\right)}\bigr)-\bar{\nabla}_{d}\bigl((\Gamma_{bce})^{\left(1\right)}+(\widetilde{\Gamma}_{bce})^{\left(1\right)}\bigr)\right).~~~~~~~~~~~~~~~~~~~~~~\label{covariantderivativeofthreeindexk2}
\end{multline}
Since the last equation is complicated we use the results given below
to get a compact form. We have

\begin{equation}
\bar{\nabla}_{c}\delta_{Y}\left(\Gamma_{bd}^{a}\right)^{\left(1\right)}-\bar{\nabla}_{d}\delta_{Y}\left(\Gamma_{bc}^{a}\right)^{\left(1\right)}=\text{\ensuremath{\mathscr{L}}}_{Y}\bar{R}^{a}\thinspace_{bcd}
\end{equation}
and from (\ref{eq:firstordergaugeinvriemann}) 

\begin{equation}
\bar{\nabla}_{c}\delta_{X}\left(\Gamma_{bd}^{e}\right)^{\left(1\right)}-\bar{\nabla}_{d}\delta_{X}\left(\Gamma_{bc}^{e}\right)^{\left(1\right)}=\text{\ensuremath{\mathscr{L}}}_{X}\bar{R}^{e}\thinspace_{bcd}=(R^{e}\thinspace_{bcd}){}^{\left(1\right)}-2\bar{\nabla}_{[c}(\widetilde{\Gamma}_{d]b}^{e})^{\left(1\right)},
\end{equation}
and 

\begin{equation}
\text{\ensuremath{\mathscr{L}}}_{X}\left(\bar{\nabla}_{c}\bigl((\Gamma_{bd}^{a}){}^{\left(1\right)}+(\widetilde{\Gamma}_{bd}^{a}){}^{\left(1\right)}\bigr)-\bar{\nabla}_{d}\bigl((\Gamma_{bc}^{a}){}^{\left(1\right)}+(\widetilde{\Gamma}_{bc}^{a}){}^{\left(1\right)}\bigr)\right)=\text{\ensuremath{\mathscr{L}}}_{X}(2(R^{a}\thinspace_{bcd}){}^{\left(1\right)}-\text{\ensuremath{\mathscr{L}}}_{X}\bar{R}^{a}\thinspace_{bcd}),
\end{equation}
and also

\begin{multline}
\text{\ensuremath{\mathscr{L}}}_{X}\overline{g}^{ea}\left(\bar{\nabla}_{c}\bigl((\Gamma_{bde})^{\left(1\right)}+(\widetilde{\Gamma}_{bde})^{\left(1\right)}\bigr)-\bar{\nabla}_{d}\bigl((\Gamma_{bce})^{\left(1\right)}+(\widetilde{\Gamma}_{bce})^{\left(1\right)}\bigr)\right)\\
=-(\bar{\nabla}^{a}X_{e}+\bar{\nabla}_{e}X^{a})\left(2(R^{e}\thinspace_{bcd}){}^{\left(1\right)}-\text{\ensuremath{\mathscr{L}}}_{X}\bar{R}^{e}\thinspace_{bcd}\right),~~~~~~~~~~~~~~~~~~~~~~~~~~~\label{equationn}
\end{multline}
and

\begin{equation}
(\Gamma_{ed}^{a}){}^{\left(1\right)}+(\widetilde{\Gamma}_{ed}^{a}){}^{\left(1\right)}-\bar{\nabla}_{d}(h_{e}^{a}+\widetilde{h}_{e}^{a})=-\left((\Gamma_{d}\thinspace^{a}\thinspace_{e}){}^{\left(1\right)}+(\widetilde{\Gamma}_{d}\thinspace^{a}\thinspace_{e}){}^{\left(1\right)}\right).
\end{equation}
Similarly we have

\begin{equation}
(\Gamma_{ec}^{a}){}^{\left(1\right)}+(\widetilde{\Gamma}_{ec}^{a}){}^{\left(1\right)}-\bar{\nabla}_{c}(h_{e}^{a}+\widetilde{h}_{e}^{a})=-\left((\Gamma_{c}\thinspace^{a}\thinspace_{e}){}^{\left(1\right)}+(\widetilde{\Gamma}_{c}\thinspace^{a}\thinspace_{e}){}^{\left(1\right)}\right)
\end{equation}
and

\begin{equation}
\delta_{X}(\Gamma_{ce}^{a}){}^{\left(1\right)}-\bar{\nabla}_{c}(\bar{\nabla}^{a}X_{e}+\bar{\nabla}_{e}X^{a})=-\delta_{X}\left(\Gamma_{c}\thinspace^{a}\thinspace_{e}\right)^{\left(1\right)}
\end{equation}
and also

\begin{equation}
\delta_{X}(\Gamma_{de}^{a}){}^{\left(1\right)}-\bar{\nabla}_{d}(\bar{\nabla}^{a}X_{e}+\bar{\nabla}_{e}X^{a})=-\delta_{X}\left(\Gamma_{d}\thinspace^{a}\thinspace_{e}\right)^{\left(1\right)}.
\end{equation}
Inserting the above results we obtain
\begin{align}
\bar{4\nabla}_{[c}K_{d]b}^{a}=\bar{2\nabla}_{[c}\widetilde{K}_{d]b}^{a}+2\text{\ensuremath{\mathscr{L}}}_{X}(R^{a}\thinspace_{bcd}){}^{\left(1\right)}-\text{\ensuremath{\mathscr{L}}}_{X}^{2}\bar{R}^{a}\thinspace_{bcd}~~~~~~~~~~~~~~~~~~~~~~~~~~~~~~~~~~~~~~~~~~~~~~~~~~\label{eqnn}\\
+(h_{e}^{a}+\widetilde{h}_{e}^{a})\text{\ensuremath{\mathscr{L}}}_{X}\bar{R}^{e}\thinspace_{bcd}+\text{\ensuremath{\mathscr{L}}}_{Y}\bar{R}^{a}\thinspace_{bcd}+(\bar{\nabla}^{a}X_{e}+\bar{\nabla}_{e}X^{a})\left(2(R^{e}\thinspace_{bcd}){}^{\left(1\right)}-\text{\ensuremath{\mathscr{L}}}_{X}\bar{R}^{e}\thinspace_{bcd}\right)\nonumber \\
-\left((\Gamma_{d}\thinspace^{a}\thinspace_{e}){}^{\left(1\right)}+(\widetilde{\Gamma}_{d}\thinspace^{a}\thinspace_{e}){}^{\left(1\right)}\right)\delta_{X}(\Gamma_{cb}^{e}){}^{\left(1\right)}+\left((\Gamma_{c}\thinspace^{a}\thinspace_{e}){}^{\left(1\right)}+(\widetilde{\Gamma}_{c}\thinspace^{a}\thinspace_{e}){}^{\left(1\right)}\right)\delta_{X}(\Gamma_{db}^{e}){}^{\left(1\right)}~~~~~~~\nonumber \\
+\left((\Gamma_{bd}^{a}){}^{\left(1\right)}+(\widetilde{\Gamma}_{bd}^{e}){}^{\left(1\right)}\right)\delta_{X}(\Gamma_{c}\thinspace^{a}\thinspace_{e})^{\left(1\right)}-\left((\Gamma_{bc}^{e}){}^{\left(1\right)}+(\widetilde{\Gamma}_{bc}^{e}){}^{\left(1\right)}\right)\delta_{X}(\Gamma_{d}\thinspace^{a}\thinspace_{e})^{\left(1\right)},~~~~~~~~~~~\nonumber 
\end{align}
which can be rewritten as
\begin{align}
\bar{4\nabla}_{[c}K_{d]b}\thinspace^{a}=\bar{2\nabla}_{[c}\widetilde{K}_{d]b}\thinspace^{a}-4\widetilde{h}_{e}^{a}\bar{\nabla}_{[c}(\widetilde{\Gamma}_{d]b}\thinspace^{e})^{\left(1\right)}+2(\widetilde{\Gamma}_{d}\thinspace^{a}\thinspace_{e})^{\left(1\right)}(\widetilde{\Gamma}_{cb}\thinspace^{e})^{\left(1\right)}-2(\widetilde{\Gamma}_{c}\thinspace^{a}\thinspace_{e})^{\left(1\right)}(\widetilde{\Gamma}_{db}\thinspace^{e})^{\left(1\right)}\nonumber \\
+2\text{\ensuremath{\mathscr{L}}}_{X}(R^{a}\thinspace_{bcd}){}^{\left(1\right)}+\left(\text{\ensuremath{\mathscr{L}}}_{Y}-\text{\ensuremath{\mathscr{L}}}_{X}^{2}\right)\bar{R}^{a}\thinspace_{bcd}+2h_{e}^{a}(R^{e}\thinspace_{bcd}){}^{\left(1\right)}~~~~~~~~~~~~~~~~~~~~~~~~\nonumber \\
+2(\Gamma_{c}\thinspace^{a}\thinspace_{e}){}^{\left(1\right)}(\Gamma_{db}\thinspace^{e}){}^{\left(1\right)}-2(\Gamma_{d}\thinspace^{a}\thinspace_{e}){}^{\left(1\right)}(\Gamma_{cb}\thinspace^{e}){}^{\left(1\right)}.~~~~~~~~~~~~~~~~~~~~~~~~~~~~~~~~~~~~~~~~\label{ddd}
\end{align}
Using the last expression we can construct the second order perturbation
of the Riemann tensor (\ref{eq:secondorderriemann} ) in terms of
gauge-invariant and variant quantities

\begin{multline}
(R^{a}\thinspace_{bcd})^{(2)}=\bar{\nabla}_{[c}\widetilde{K}_{d]b}^{a}-2\widetilde{h}_{e}^{a}\bar{\nabla}_{[c}(\widetilde{\Gamma}_{d]b}^{e})^{\left(1\right)}+(\widetilde{\Gamma}_{d}\thinspace^{a}\thinspace_{e})^{\left(1\right)}(\widetilde{\Gamma}_{cb}^{e})^{\left(1\right)}-(\widetilde{\Gamma}_{c}\thinspace^{a}\thinspace_{e})^{\left(1\right)}(\widetilde{\Gamma}_{db}^{e})^{\left(1\right)}\\
+\text{\ensuremath{\mathscr{L}}}_{X}(R^{a}\thinspace_{bcd}){}^{\left(1\right)}+\frac{1}{2}\left(\text{\ensuremath{\mathscr{L}}}_{Y}-\text{\ensuremath{\mathscr{L}}}_{X}^{2}\right)\bar{R}^{a}\thinspace_{bcd},~~~~~~~~~~~~~~~~~~~~~~~~~~~~~~~~~~~~~~~~~~~~~~~~~\label{gaugeinvarianttheoryriemann}
\end{multline}
where the second line shows the gauge-variant terms and this result
is consistent with the aim of the gauge-invariant perturbation theory.
Contraction of the indices yields the decomposition of the second
order Ricci tensor

\begin{multline}
(R_{ab})^{(2)}=\bar{\nabla}_{[c}\widetilde{K}_{a]b}^{c}-2\widetilde{h}_{e}^{c}\bar{\nabla}_{[c}(\widetilde{\Gamma}_{a]b}^{e})^{\left(1\right)}+(\widetilde{\Gamma}_{a}\thinspace^{c}\thinspace_{e})^{\left(1\right)}(\widetilde{\Gamma}_{cb}^{e})^{\left(1\right)}-(\widetilde{\Gamma}_{c}\thinspace^{c}\thinspace_{e})^{\left(1\right)}(\widetilde{\Gamma}_{ab}^{e})^{\left(1\right)}\\
+\text{\ensuremath{\mathscr{L}}}_{X}(R{}_{ab}){}^{\left(1\right)}+\frac{1}{2}\left(\text{\ensuremath{\mathscr{L}}}_{Y}-\text{\ensuremath{\mathscr{L}}}_{X}^{2}\right)\bar{R}{}_{ab}.~~~~~~~~~~~~~~~~~~~~~~~~~~~~~~~~~~~~~~~~~~~~~~~~~~~~~~~~~\label{gaugeinvariantricci}
\end{multline}
The second order Ricci scalar (\ref{eq:secondorderscalarcurvature})
becomes

\begin{multline}
(R)^{(2)}=\bar{\nabla}_{[c}\widetilde{K}_{a]}{}^{ac}-2\widetilde{h}_{e}^{c}\bar{\nabla}_{[c}(\widetilde{\Gamma}_{a]}{}^{ae})^{\left(1\right)}+2(\widetilde{\Gamma}_{[c}\thinspace^{ae})^{\left(1\right)}(\widetilde{\Gamma}_{a]}\thinspace^{c}\thinspace_{e})^{\left(1\right)}+\overline{g}^{ab}\text{\ensuremath{\mathscr{L}}}_{X}(R{}_{ab}){}^{\left(1\right)}\\
+\frac{1}{2}\overline{g}^{ab}(\text{\ensuremath{\mathscr{L}}}_{Y}-\text{\ensuremath{\mathscr{L}}}_{X}^{2})\bar{R}{}_{ab}-(\widetilde{h}^{ab}-\text{\ensuremath{\mathscr{L}}}_{X}\overline{g}^{ab})\left(2\bar{\nabla}_{[c}\widetilde{\Gamma}_{a]b}^{c}+\text{\ensuremath{\mathscr{L}}}_{X}\bar{R}_{ab}\right)~~~~~~~~~~~~~~~\\
+(\widetilde{h}^{ac}-\text{\ensuremath{\mathscr{L}}}_{X}\overline{g}^{ac})\left(\widetilde{h}_{cb}+\text{\ensuremath{\mathscr{L}}}_{X}\overline{g}_{cb}\right)\bar{R}{}_{a}^{b}-\frac{1}{2}\left(\widetilde{k}_{ab}+2\text{\ensuremath{\mathscr{L}}}_{X}h_{ab}+(\text{\ensuremath{\mathscr{L}}}_{Y}-\text{\ensuremath{\mathscr{L}}}_{X}^{2})\overline{g}_{ab}\right)\bar{R}^{ab},\label{eee}
\end{multline}
which reduces to
\begin{align}
(R)^{(2)}=\bar{\nabla}_{[c}\widetilde{K}_{a]}{}^{ac}-2\widetilde{h}_{e}^{c}\bar{\nabla}_{[c}\widetilde{\Gamma}_{a]}{}^{ae}+2\widetilde{\Gamma}_{[c}\thinspace^{ae}\widetilde{\Gamma}_{a]}\thinspace^{c}\thinspace_{e}-2\widetilde{h}^{ab}\bar{\nabla}_{[c}\widetilde{\Gamma}_{a]b}^{c}-\frac{1}{2}\widetilde{k}_{ab}\bar{R}^{ab}+\widetilde{h}^{ac}\widetilde{h}_{bc}\bar{R}{}_{c}^{b}~~~~~~~~\label{ff}\\
+\overline{g}^{ab}\text{\ensuremath{\mathscr{L}}}_{X}(R{}_{ab}){}^{\left(1\right)}+\frac{1}{2}\overline{g}^{ab}\bigl(\text{\ensuremath{\mathscr{L}}}_{Y}-\text{\ensuremath{\mathscr{L}}}_{X}^{2}\bigr)\bar{R}{}_{ab}-\bar{R}^{ab}\text{\ensuremath{\mathscr{L}}}_{X}h_{ab}-\frac{1}{2}\bar{R}^{ab}\bigl(\text{\ensuremath{\mathscr{L}}}_{Y}-\text{\ensuremath{\mathscr{L}}}_{X}^{2}\bigr)\overline{g}_{ab}~~~~~\nonumber \\
-\widetilde{h}^{ab}\text{\ensuremath{\mathscr{L}}}_{X}\bar{R}_{ab}+(R{}_{ab}){}^{\left(1\right)}\text{\ensuremath{\mathscr{L}}}_{X}\overline{g}^{ab}+\widetilde{h}^{ac}\bar{R}{}_{a}^{b}\text{\ensuremath{\mathscr{L}}}_{X}\overline{g}_{cb}-\widetilde{h}_{cb}\bar{R}{}_{a}^{b}\text{\ensuremath{\mathscr{L}}}_{X}\overline{g}^{ac}-\bar{R}{}_{a}^{b}\text{\ensuremath{\mathscr{L}}}_{X}\overline{g}^{ac}\text{\ensuremath{\mathscr{L}}}_{X}\overline{g}_{cb}.\nonumber 
\end{align}
Let us concentrate on the gauge variant terms: we can write

\begin{equation}
\overline{g}^{ab}\text{\ensuremath{\mathscr{L}}}_{X}(R{}_{ab}){}^{\left(1\right)}+(R_{ab}){}^{\left(1\right)}\text{\ensuremath{\mathscr{L}}}_{X}\overline{g}^{ab}-\bar{R}^{ab}\text{\ensuremath{\mathscr{L}}}_{X}h_{ab}=\text{\ensuremath{\mathscr{L}}}_{X}(R){}^{\left(1\right)}+h_{ab}\text{\ensuremath{\mathscr{L}}}_{X}\bar{R}^{ab},
\end{equation}
and 

\begin{multline}
\frac{1}{2}\overline{g}^{ab}(\text{\ensuremath{\mathscr{L}}}_{Y}-\text{\ensuremath{\mathscr{L}}}_{X}^{2})\bar{R}{}_{ab}-\widetilde{h}^{ab}\text{\ensuremath{\mathscr{L}}}_{X}\bar{R}_{ab}-\frac{1}{2}\bar{R}^{ab}(\text{\ensuremath{\mathscr{L}}}_{Y}-\text{\ensuremath{\mathscr{L}}}_{X}^{2})\overline{g}_{ab}\\
=\frac{1}{2}(\text{\ensuremath{\mathscr{L}}}_{Y}-\text{\ensuremath{\mathscr{L}}}_{X}^{2})\bar{R}-h_{ab}\text{\ensuremath{\mathscr{L}}}_{X}\bar{R}^{ab}-2h^{ab}\bar{R}{}_{a}^{d}\text{\ensuremath{\mathscr{L}}}_{X}\bar{g}_{db}-\bar{R}{}_{d}^{a}\text{\ensuremath{\mathscr{L}}}_{X}\bar{g}_{ca}\text{\ensuremath{\mathscr{L}}}_{X}\overline{g}^{dc},\label{bd}
\end{multline}
and also
\begin{equation}
\widetilde{h}^{ac}\bar{R}{}_{a}^{b}\text{\ensuremath{\mathscr{L}}}_{X}\overline{g}_{cb}-\widetilde{h}_{cb}\bar{R}{}_{a}^{b}\text{\ensuremath{\mathscr{L}}}_{X}\overline{g}^{ac}-\bar{R}{}_{a}^{b}\text{\ensuremath{\mathscr{L}}}_{X}\overline{g}^{ac}\text{\ensuremath{\mathscr{L}}}_{X}\overline{g}_{cb}=-\widetilde{h}_{cb}\bar{R}{}_{a}^{b}\text{\ensuremath{\mathscr{L}}}_{X}\overline{g}^{ac}+h^{ac}\bar{R}{}_{a}^{b}\text{\ensuremath{\mathscr{L}}}_{X}\overline{g}_{cb}.
\end{equation}
Finally the second order scalar curvature yields

\begin{multline}
(R)^{(2)}=\bar{\nabla}_{[c}\widetilde{K}_{a]}{}^{ac}-2\widetilde{h}_{e}^{c}\bar{\nabla}_{[c}\widetilde{\Gamma}_{a]}{}^{ae}+\widetilde{\Gamma}_{[c}\thinspace^{ae}\widetilde{\Gamma}_{a]}\thinspace^{c}\thinspace_{e}-2\widetilde{h}^{ab}\bar{\nabla}_{[c}\widetilde{\Gamma}_{a]b}^{c}-\frac{1}{2}\bar{R}^{ab}\bigl(\widetilde{k}_{ab}-\widetilde{h}_{a}^{c}\widetilde{h}_{bc}\bigr)\\
+\text{\ensuremath{\mathscr{L}}}_{X}(R){}^{\left(1\right)}+\frac{1}{2}(\text{\ensuremath{\mathscr{L}}}_{Y}-\text{\ensuremath{\mathscr{L}}}_{X}^{2})\bar{R}.~~~~~~~~~~~~~~~~~~~~~~~~~~~~~~~~~~~~~~~~~~~~~~~~~~~~~~~~~~~~~~~~~~~\label{gaugeinvariantscalarcurvature}
\end{multline}
Now we can compute the second order perturbation of the cosmological
Einstein tensor (\ref{eq:secondordercosmologicaleinstein}) in terms
of gauge-variant and invariant quantities. From the previous results
we get

\begin{multline}
({\cal {G}}_{ab})^{(2)}=\bar{\nabla}_{[c}\widetilde{K}_{a]b}^{c}-2\widetilde{h}_{e}\thinspace^{c}\bar{\nabla}_{[c}\widetilde{\Gamma}_{a]b}^{e}+2\widetilde{\Gamma}_{b[c}^{e}\widetilde{\Gamma}_{a]}\thinspace^{c}{}_{e}+\text{\ensuremath{\mathscr{L}}}_{X}(R{}_{ab}){}^{\left(1\right)}+\frac{1}{2}(\text{\ensuremath{\mathscr{L}}}_{Y}-\text{\ensuremath{\mathscr{L}}}_{X}^{2})\bar{R}{}_{ab}\\
-\frac{1}{2}\bar{g}_{ab}\Biggl(\bar{\nabla}_{[c}\widetilde{K}_{d]}{}^{dc}-2\widetilde{h}_{e}^{c}\bar{\nabla}_{[c}\widetilde{\Gamma}_{d]}\thinspace^{de}+\bar{R}^{cd}\bigl(\widetilde{h}^{e}\thinspace_{d}\widetilde{h}_{ce}-\frac{1}{2}\widetilde{k}_{cd}\bigr)~~~~~~~~~~~~~~~~~~~~~~~~~~~~~~~~\\
+2\widetilde{\Gamma}^{d}\thinspace_{[c}\thinspace^{e}\widetilde{\Gamma}_{d]}\thinspace^{c}{}_{e}-2\widetilde{h}^{cd}\bar{\nabla}_{[e}\widetilde{\Gamma}_{c]d}^{e}+\text{\ensuremath{\mathscr{L}}}_{X}(R){}^{\left(1\right)}+\frac{1}{2}(\text{\ensuremath{\mathscr{L}}}_{Y}-\text{\ensuremath{\mathscr{L}}}_{X}^{2})\bar{R}\Biggr)~~~~~~~~~~~~~~~~~~\\
-\frac{1}{2}\left(\widetilde{h}_{ab}+\text{\ensuremath{\mathscr{L}}}_{X}\bar{g}_{ab}\right)\left(2\bar{\nabla}_{[c}\widetilde{\Gamma}_{d]}{}^{dc}-\bar{R}_{dc}\widetilde{h}^{dc}+\text{\ensuremath{\mathscr{L}}}_{X}\bar{R}\right)~~~~~~~~~~~~~~~~~~~~~~~~~~~~~~~~~~~\\
+\left(\widetilde{k}_{ab}+2\text{\ensuremath{\mathscr{L}}}_{X}h_{ab}+(\text{\ensuremath{\mathscr{L}}}_{Y}-\text{\ensuremath{\mathscr{L}}}_{X}^{2})\overline{g}{}_{ab}\right)\bigl(\frac{\Lambda}{2}-\frac{\bar{R}}{4}\bigr),~~~~~~~~~~~~~~~~~~~~~~~~~~~~~~~~~~~~~~~~~~~\label{DSAASW}
\end{multline}
which reduces to
\begin{multline}
({\cal {G}}_{ab})^{(2)}=\bar{\nabla}_{[c}\widetilde{K}_{a]b}^{c}-2\widetilde{h}_{e}\thinspace^{c}\bar{\nabla}_{[c}\widetilde{\Gamma}_{a]b}^{e}+2\widetilde{\Gamma}_{b[c}^{e}\widetilde{\Gamma}_{a]}\thinspace^{c}{}_{e}+\widetilde{k}_{ab}\bigl(\frac{\Lambda}{2}-\frac{\bar{R}}{4}\bigr)-\frac{1}{2}\widetilde{h}_{ab}\bigl(2\bar{\nabla}_{[c}\widetilde{\Gamma}_{d]}{}^{dc}-\bar{R}_{dc}\widetilde{h}^{dc}\bigr)~~~~\\
-\frac{1}{2}\bar{g}_{ab}\Biggl(\bar{\nabla}_{[c}\widetilde{K}_{d]}{}^{dc}-2\widetilde{h}_{e}^{c}\bar{\nabla}_{[c}\widetilde{\Gamma}_{d]}\thinspace^{de}+\bar{R}^{cd}\bigl(\widetilde{h}^{e}\thinspace_{d}\widetilde{h}_{ce}-\frac{1}{2}\widetilde{k}_{cd}\bigr)+2\widetilde{\Gamma}^{d}\thinspace_{[c}\thinspace^{e}\widetilde{\Gamma}_{d]}\thinspace^{c}{}_{e}-2\widetilde{h}^{cd}\bar{\nabla}_{[e}\widetilde{\Gamma}_{c]d}^{e}\Biggr)\\
+\text{\ensuremath{\mathscr{L}}}_{X}(R{}_{ab}){}^{\left(1\right)}-\frac{1}{2}(R)^{\left(1\right)}\text{\ensuremath{\mathscr{L}}}_{X}\bar{g}_{ab}+\bigl(\Lambda-\frac{\bar{R}}{2}\bigr)\text{\ensuremath{\mathscr{L}}}_{X}h_{ab}-\frac{1}{2}\widetilde{h}_{ab}\text{\ensuremath{\mathscr{L}}}_{X}\bar{R}-\frac{1}{2}\bar{g}_{ab}\text{\ensuremath{\mathscr{L}}}_{X}(R){}^{\left(1\right)}\\
-\frac{1}{4}\bar{g}_{ab}(\text{\ensuremath{\mathscr{L}}}_{Y}-\text{\ensuremath{\mathscr{L}}}_{X}^{2})\bar{R}+\frac{1}{2}(\text{\ensuremath{\mathscr{L}}}_{Y}-\text{\ensuremath{\mathscr{L}}}_{X}^{2})\overline{R}{}_{ab}+\bigl(\frac{\Lambda}{2}-\frac{\bar{R}}{4}\bigr)(\text{\ensuremath{\mathscr{L}}}_{Y}-\text{\ensuremath{\mathscr{L}}}_{X}^{2})\overline{g}{}_{ab},~~~~~~~~~~~~~~~~~~~~\label{sasddes}
\end{multline}
where the first two lines denote the gauge-invariant part. Let us
consider the gauge-variant terms. We can collect the third line as

\begin{multline}
\text{\ensuremath{\mathscr{L}}}_{X}(R{}_{ab}){}^{\left(1\right)}-\frac{1}{2}(R)^{\left(1\right)}\text{\ensuremath{\mathscr{L}}}_{X}\bar{g}_{ab}+\bigl(\Lambda-\frac{\bar{R}}{2}\bigr)\text{\ensuremath{\mathscr{L}}}_{X}h_{ab}-\frac{1}{2}\widetilde{h}_{ab}\text{\ensuremath{\mathscr{L}}}_{X}\bar{R}-\frac{1}{2}\bar{g}_{ab}\text{\ensuremath{\mathscr{L}}}_{X}(R){}^{\left(1\right)}\\
=\text{\ensuremath{\mathscr{L}}}_{X}({\cal {G}}_{ab}){}^{\left(1\right)}+\frac{1}{2}\text{\ensuremath{\mathscr{L}}}_{X}\bar{g}_{ab}\text{\ensuremath{\mathscr{L}}}_{X}\bar{R}~~~~~~~~~~~~~~~~~~\label{dsedsd}
\end{multline}
and the terms on the last line yield

\begin{multline}
-\frac{1}{4}\bar{g}_{ab}\left(\text{\ensuremath{\mathscr{L}}}_{Y}-\text{\ensuremath{\mathscr{L}}}_{X}^{2}\right)\bar{R}+\frac{1}{2}(\text{\ensuremath{\mathscr{L}}}_{Y}-\text{\ensuremath{\mathscr{L}}}_{X}^{2})\overline{R}{}_{ab}+\bigl(\frac{\Lambda}{2}-\frac{\bar{R}}{4}\bigr)(\text{\ensuremath{\mathscr{L}}}_{Y}-\text{\ensuremath{\mathscr{L}}}_{X}^{2})\overline{g}{}_{ab}\\
=\frac{1}{2}\bigl(\text{\ensuremath{\mathscr{L}}}_{Y}-\text{\ensuremath{\mathscr{L}}}_{X}^{2}\bigr)\bar{{\cal {G}}}{}_{ab}-\frac{1}{2}\text{\ensuremath{\mathscr{L}}}_{X}\bar{R}\text{\ensuremath{\mathscr{L}}}_{X}\bar{g}_{ab}.~~~~~~~~~~~~~~~~~~~~~~~~~~~~~~\label{sdesasd}
\end{multline}
Finally we obtain the second order cosmological Einstein tensor 

\begin{multline}
({\cal {G}}_{ab})^{(2)}=\bar{\nabla}_{[c}\widetilde{K}_{a]b}^{c}-2\widetilde{h}_{e}\thinspace^{c}\bar{\nabla}_{[c}\widetilde{\Gamma}_{a]b}^{e}+2\widetilde{\Gamma}_{b[c}^{e}\widetilde{\Gamma}_{a]}\thinspace^{c}{}_{e}+\widetilde{k}_{ab}\bigl(\frac{\Lambda}{2}-\frac{\bar{R}}{4}\bigr)-\frac{1}{2}\widetilde{h}_{ab}\bigl(2\bar{\nabla}_{[c}\widetilde{\Gamma}_{d]}{}^{dc}-\bar{R}_{dc}\widetilde{h}^{dc}\bigr)\\
-\frac{1}{2}\bar{g}_{ab}\Biggl(\bar{\nabla}_{[c}\widetilde{K}_{d]}{}^{dc}-2\widetilde{h}_{e}^{c}\bar{\nabla}_{[c}\widetilde{\Gamma}_{d]}\thinspace^{de}+\bar{R}^{cd}\left(\widetilde{h}^{e}\thinspace_{d}\widetilde{h}_{ce}-\frac{1}{2}\widetilde{k}_{cd}\right)+2\widetilde{\Gamma}^{d}\thinspace_{[c}\thinspace^{e}\widetilde{\Gamma}_{d]}\thinspace^{c}{}_{e}-2\widetilde{h}^{cd}\bar{\nabla}_{[e}\widetilde{\Gamma}_{c]d}^{e}\Biggr)\\
+\text{\ensuremath{\mathscr{L}}}_{X}({\cal {G}}{}_{ab}){}^{\left(1\right)}+\frac{1}{2}\left(\text{\ensuremath{\mathscr{L}}}_{Y}-\text{\ensuremath{\mathscr{L}}}_{X}^{2}\right)\bar{{\cal {G}}}{}_{ab},~~~~~~~~~~~~~~~~~~~~~~~~~~~~~~~~~~~~~~~~~~~~~~~~~~~~~~~~~~~~\label{eq:secondordercosmoeinstein}
\end{multline}
where the gauge-variant terms vanish when $\bar{g}$ is solution to
the background equations and $h$ is a solution of the first order
linearized equations. In this case we arrive at a pure gauge-invariant
second order cosmological Einstein tensor.
\begin{acknowledgments}
This work was done in the Physics Department of the Middle East Technical
University. The Author would like to thank Prof. Dr. Bayram Tekin
for his comments and extended discussions on conserved charges in
cosmological Einstein gravity.
\end{acknowledgments}

\end{document}